\documentclass[conference]{IEEEtran}
\IEEEoverridecommandlockouts
\usepackage{float}
\usepackage{cite}
\usepackage{amsmath,amssymb,amsfonts}
\usepackage{algorithmic}
\usepackage{graphicx}
\usepackage{caption}
\usepackage{textcomp}
\usepackage{xcolor}
\usepackage{bm}
\usepackage{amsmath}
\usepackage{amssymb}
\usepackage{enumitem}
\usepackage[bookmarks]{hyperref}
\def\BibTeX{{\rm B\kern-.05em{\sc i\kern-.025em b}\kern-.08em
    T\kern-.1667em\lower.7ex\hbox{E}\kern-.125emX}}
\begin{document}

\captionsetup[table]{
    position=above,        
    justification=centering,  
    font=normal,             
    labelfont=normal,        
    textfont=normal,         
    labelsep=colon          
}

\title{Physics Meets Pixels: \\
PDE Models in Image Processing
}

\author{
    \IEEEauthorblockN{Alejandro Garnung Menéndez, M.Sc., PhD Candidate\textsuperscript{1}}
    \IEEEauthorblockA{
        \textsuperscript{1}\textit{University of Oviedo, Gijón Polytechnic School of Engineering, Asturias, España} \\
        \texttt{uo269564@uniovi.es}, \texttt{garnungalejandro@gmail.com}
    }
}

\maketitle

\begin{abstract}
Partial Differential Equations (PDEs) have long been recognized as powerful tools for image processing and analysis, providing a framework to model and exploit structural and geometric properties inherent in visual data. Over the years, numerous PDE-based models have been developed and refined, inspired by natural analogies between physical phenomena and image spaces. These methods have proven highly effective in a wide range of applications, including denoising, deblurring, sharpening, inpainting, feature extraction, and others.
This work provides a theoretical and computational exploration of both fundamental and innovative PDE models applied to image processing, accompanied by extensive numerical experimentation and objective and subjective analysis. Building upon well-established techniques, we introduce novel physical-based PDE models specifically designed for various image processing tasks. These models incorporate mathematical principles and approaches that, to the best of our knowledge, have not been previously applied in this domain, showcasing their potential to address challenges beyond the capabilities of traditional and existing PDE methods.

By formulating and solving these mathematical models, we demonstrate their effectiveness in advancing image processing tasks while retaining a rigorous connection to their theoretical underpinnings. This work seeks to bridge foundational concepts and cutting-edge innovations, contributing to the evolution of PDE methodologies in digital image processing and related interdisciplinary fields.
\end{abstract}


\section{Introduction}
\IEEEPARstart{P}{DE}-based methods, beginning with the foundational heat equation, provide a dynamic and interpretable framework for image processing, standing in contrast to the opaque nature of many “black box” algorithms. These methods are particularly effective for a wide range of tasks like denoising, edge detection, segmentation, inpainting, enhancement/sharpening, deblurring/deconvolution and many others. As mentioned in \cite{Gary}, an image processing system driven by PDEs allows visual feedback, enabling users to adjust the process and see real-time changes, making the approach more transparent and adaptable, ideal for applications like image restoration, segmentation, and video frame fusion.
Over the years, it has become clear that PDEs are a highly effective way to smooth images. When the PDE is linear, such as the classic heat equation, it can be viewed as a convolution operator; however, this approach does not preserve edges well. In contrast, when higher-order derivatives or nonlinear terms are introduced—like those in anisotropic diffusion—PDE-based models surpass traditional filtering techniques, offering more natural image processing and improved restoration capabilities.
One key model is the Perona-Malik (P-M) model, which modifies the heat equation to allow for edge-preserving smoothing, weighting the gradient magnitude by a diffusion edge-stopping function that decreases correlatively, thus reducing smoothing at edges. As noted in the non-local mean (NLM) filtering formulation \cite{Buades2005}, which processes patches of pixels for more accurate denoising, the bilateral filter \cite{Tomasi} is closely related to the P-M model, as it operates both spatially and on the intensity domain and applies a similar concept of edge preservation, but in a local, non-iterative, and computationally simpler way. Moreover, it is well known that Gaussian filtering is equivalent to the heat equation. Both methods provide a scale-space perspective for noise filtering, which allows for better handling of fine details and. However, although Gaussian filtering can be implemented through convolution via the Fourier transform in the frequency domain, it is the PDE that offers a clearer intuition and spatial correspondence and localization, something that even the frequency-based approach of the traditional method lacks.
Like these, many other examples show that PDEs offer several advantages over traditional simple approaches in image processing due to their solid theoretical foundation and their ability to model complex processes naturally. PDEs are rooted in well-established mathematical principles from physics, such as heat diffusion and wave propagation. This continuous framework provides a more intuitive understanding of how to process and transform images while preserving important features like edges. Since PDEs describe processes over a continuous domain, they provide a more flexible and adaptable approach compared to traditional methods. The theory behind PDEs ensures that solutions are well-defined, and once the solution’s existence and uniqueness are proven, they can be discretized for numerical applications.
Additionally, PDEs allow for the incorporation of more complex terms, such as anisotropic diffusion and nonlinearity or high-order derivatives to preserve discontinuities. Also, the integration of PDEs with variational methods has led to immeasurable advancements in image processing. By combining these two approaches, we can effectively model physical processes and optimize solutions to complex problems. PDEs, when applied to image processing, are used to smooth images, while variational methods introduce regularization techniques that constrain the solution space and stabilize terribly ill-posed problems. Together, these methodologies enable the extension of existing models to address more specialized tasks, offering a robust and versatile framework for advanced image analysis.
It is indisputable that PDEs have a wide range of applications in image processing and computer vision. This work provides a review and experimentation with various PDE models, both classical and innovative, applied to typical image processing tasks. The goal is to offer a qualitative and quantitative analysis of these models, highlighting the underlying theory of physical PDEs and their analogies to image processing. The emphasis will be put on analyzing results obtained from different approaches to specific tasks in image processing. Additionally, physical models applied in innovative ways to images are proposed, to enrich the existing literature in this field.
    
\section{Related work}
Hewer and Lau \cite{Gary} mention one of the first usage of nonlinear PDEs applied to image processing, initially borrowed from computational fluid dynamics (CFD) to capture image discontinuities at edges. These methods, rooted in variational principles, involve two key steps: formulating a relevant variational problem and efficiently solving it. The heat equation, for example, is derived from this variational framework, allowing for iterative image processing. As time progresses, the image evolves by minimizing the energy functional, enhancing details and edges while reducing noise.
In his seminal work, Weickert introduced a method to improve image diffusion by considering local image structures. Nonlinear diffusion, in this context, is directional and adapts to the image’s features. A limitation of earlier models was the local estimation of the diffusion direction; he addressed this by using the structure tensor, which captures image gradients and their interactions, allowing for more accurate diffusion that follows the image’s directional features. This method enhances image processing by guiding the diffusion along edges and within coherent regions, leading to more effective, context-aware image processing.
Recently, a new approach to denoising tourist street view images using PDEs is proposed in  \cite{Yang2021}, using an adaptive total variation (TV) diffusion model to improve image quality by removing noise while preserving edge and texture details. The model enhances the classic TV method and the Gaussian thermal diffusion model. By modifying the diffusion coefficient and introducing a curvature operator, the model effectively reduces the staircase and outperforms traditional denoising techniques like the ROF (Rudin-Osher-Fatemi) model. Experimental results show that the proposed model offers better denoising performance and faster execution than traditional methods, especially in preserving edge information. 
In \cite{Wang2022}, the authors use physical PDEs in nonlinear diffusion filtering to enhance image contrast. By adding a velocity term to the diffusion equation, they achieve smoother diffusion that adapts to the image’s features. They incorporate a coupled denoising model to preserve details while reducing noise. The approach is grounded in multi-physics field coupling theory, which helps denoise the interferogram while maintaining edges and geometric structures. Additionally, anisotropic diffusion guides the process based on local image structures, improving contrast without distortion and enhancing image quality effectively and efficiently.
The authors in \cite{Yu2021} apply PDEs to enhance video surveillance images in cross-border e-commerce logistics. By addressing limitations of traditional methods, the authors propose a hybrid approach combining PDEs with histogram equalization. Key innovations include a gradient transformation function to enhance weak textures and a regression-based formula for adaptive fractional derivative orders, reducing manual tuning. The method improves clarity, texture details, brightness, and contrast without causing color distortion, offering an efficient solution for challenging surveillance environments. Similarly, \cite{Pan2021} highlights the use of anisotropic partial differential PDEs for automating the recognition and management of financial invoices, improving recognition accuracy and efficiency by overcoming traditional OCR limitations, and providing high-performance denoising and edge protection in image processing. 
Also, \cite{Yu2021_bis} proposes a multifeature image retrieval method based on PDEs, simulating human visual perception to analyze images, addressing the issue of indexing imbalance in multifeature image retrieval. The method introduces a query pruning algorithm that enhances retrieval speed while maintaining accuracy. Additionally, a data-dependent PDE algorithm is applied to distribute datasets across operation nodes for better parallelization, utilizing a clustering method with constraints to improve data categorization.
In \cite{Li2021_1}, the authors address image restoration issues caused by network packet loss or pixel damage, proposing a dual discriminator generative adversarial network (GAN) model that improves restoration accuracy by adding local discriminators for tracking missing regions, introducing a PDE-based restoration model which integrates a classifier and feature extraction network into the GAN framework, enforcing constraints on category, style, and content. The approach encodes physical information into the loss function of a deep neural network, to optimally solve the problem, using automatic differentiation to construct the loss function. However, the usage of pure PDE-based methods ensures physical consistency and accuracy by directly modeling real-world phenomena, while learning or purely data-driven models may lack this inherent reliability and require extensive samples to generalize effectively.

\section{Physical models for image processing}
Just as classic diffusion models such as ROF or Perona-Malik were inspired by basic physical behaviors modeled by well-known differential equations, new approaches in this field may prove promising in opening new doors to innovative solutions to tackle numerous signal or image processing tasks. In the following, we are going to provide some insight into this well-established field and present various new models that, to the best of our knowledge, have not yet been applied to image processing.

\subsection{Heat Equation}
The theory of heat equation was first developed by Joseph Fourier in 1822 through Fourier’s law, which relates the thermal current density (or heat flux vector per unit area) and the temperature gradient:
\[
\mathbf{j}=-\sigma \cdot \nabla T,
\]
where \(\sigma \) is the thermal conductivity, which is generally a tensor or reduces to a diagonal matrix or constant term with identical diagonal elements \({{\sigma }_{i}}\) for an isotropic material.

Just as heat diffuses through a material until the temperature becomes uniform, an analogy is drawn to image processing, in the sense that noise can be spread out to achieve smoother transitions in pixel intensity, resulting in a more natural and less noisy image. For the energy to be conserved, the amount of heat flowing out a given point has to be equal to the rate of change (decrease) of heat contained at that point, i.e.:
\[
\nabla \cdot \mathbf{j} = -\frac{\partial T}{\partial t}. 
\]

The combination of Fourier’s law and the conservation of energy (continuity equation) yields the heat equation. Focusing our discussion on the processing of 2D images represented on a Cartesian grid, it can be expressed as:
\[
\frac{\partial u}{\partial t} = \alpha \cdot \Delta u,
\]
with initial and boundary conditions \(u(x,y,0) = u_0\) and \(\frac{\partial u}{\partial \mathbf{n}} = 0\), where \(\Delta u \equiv \nabla^2 u \equiv \nabla \cdot (\nabla u) = \frac{\partial^2 u}{\partial x^2} + \frac{\partial^2 u}{\partial y^2}\) is the Laplacian of \(u\), representing the sum of the second partial derivatives in the spatial directions \(x\) and \(y\). Here, \(u\) is a function representing temperature (or image intensity) at each point in the spatial domain of the image \(u: \Omega \subset \mathbb{R}^2 \times \mathbb{R} \to \mathbb{R}\) (typically a bounded subset of \(\mathbb{R}^2\)), and \(\alpha \in \mathbb{R}\) is the thermal diffusivity coefficient, which determines the rate of heat flow (or intensity diffusion). PDE models evolve the image over space and time, based on underlying physical principles, so from now on the image can also be thought of as a mapping from the space-time domain to the set of possible intensity values.

This model leverages the heat equation by treating the image as a dynamic system that evolves over time. To solve this PDE, numerical methods like forward and backward differencing are employed, allowing efficient computation and precise control over the denoising process. In practice, we typically adopt Neumann boundary conditions to solve the PDE, ensuring that the boundaries of the image do not behave as sources or sinks of energy (intensity), i.e., \(\frac{\partial u}{\partial \mathbf{n}} = 0\) in \(\partial \Omega\). This approach conserves energy within the image, maintaining a zero net heat flow across the boundaries, which is necessary to ensure that the image maintains its energy value throughout the evolution.

It is worth mentioning that the heat equation is equivalent to a Gaussian blur because both describe diffusion, that is, the spreading and smoothing of information, such as heat or intensity, over time or space, governed by similar mathematical principles. With proper boundary conditions, in steady state, the heat equation provides the mean intensity value of the image. 

This is a parabolic-type, second-order, linear PDE (since \(\alpha\) does not depend on \(u\) but may vary spatially and temporally), with constant coefficients, and typically homogeneous.

\subsection{Laplace Equation}
The Laplace equation models the equilibrium of potential or scalar fields (such as electrostatic or temperature) under stationary conditions. It can be represented by:
\[
\Delta u = 0, \frac{\partial u}{\partial \mathbf{n}} = 0, 
\]
where, in the context of image processing, \(u_0\) is the initial image, e.g., the noisy image in a denoising process, and we adopt Neumann boundary conditions.  

So, it is a particularization of the heat equation in which there is no source, e.g., the steady-state electrostatics, for a volume in a free space that does not contain a charge. 

This is an elliptic-type, second-order, linear PDE, with constant coefficients, and typically homogeneous.

\subsection{Poisson Equation}
Poisson’s equation generalizes Laplace’s equation, as it is used to model potentials or scalar fields with sources or sinks (i.e. positive or negative sources) of energy, such as electrostatic potentials or gravitational fields with position-dependent charge or mass density distributions, respectively. Hence, it is the steady-state heat equation for a volume that contains a heat source. It can be defined by:
\[
\Delta u=f(x,y), \frac{\partial u}{\partial \mathbf{n}}=0,
\]
where \(u=u(x,y)\) is the potential field and \(f(x,y)\) is the source function, which can represent various quantities depending on the specific application, such as known data in a corrupted image, gradient information, or other encoded features like edges or intensity variations that guide the evolution of the image with a specific purpose.

This is an elliptic-type, second-order, linear, with either constant or variable coefficients, and generally non-homogeneous PDE.

\subsection{Diffusion Equation}
The diffusion equation is fundamental for describing the macroscopic behavior of numerous micro-particles that form a substance. It models how concentrations evolve under diffusion processes driven by Brownian motion, which arises from the random movements and collisions of particles at a microscopic scale. The equation, which is equivalent to Fick’s second law (prediction of change in concentration gradient with time due to diffusion), is usually written as:
\[
\frac{\partial u(x,y,t)}{\partial t} - \nabla \cdot \left( D(x,y) \nabla u(x,y,t) \right) = S(x,y,t), 
\]
for the concentration \(u(x,y,t)\) at spatial positions \(x\) and \(y\), and time \(t\), under the given source \(S(x,y,t)\). Here, \(D(x,y)\) represents the diffusion coefficient at position \((x,y)\). This is a typical parabolic-type, second-order, linear or nonlinear (depending on the implementation), with variable coefficients (anisotropic diffusivity) and homogeneous or non-homogeneous PDE.  

In isotropic diffusion, the diffusion coefficient \(D(x,y) = D_0 \in \mathbb{R}\) is constant and does not depend on spatial coordinates or gradients. This results in a uniform spreading of the quantity \(u(x,y,t)\) in all directions. The governing equation becomes:
\[
\frac{\partial u}{\partial t} = \nabla \cdot \left( D_0 \nabla u \right).
\]
For image processing, isotropic diffusion effectively reduces noise but also blurs edges and fine details because it treats all regions equally. This limitation is a well-known drawback of this type of PDE.

In anisotropic diffusion, the diffusion of the physical quantity is directed by the gradient of the image, allowing for greater diffusion in areas with low gradient (smooth regions) and lower diffusion in regions with high gradient (edges):  
\[
\frac{\partial u}{\partial t} = \nabla \cdot (D \nabla u) = \nabla D \cdot \nabla u + D \Delta u,  
\]
where \(D = D(x,y,t)\) is the diffusion coefficient, which may be variable in space and time.  

Thus, in image processing, anisotropic diffusion-based approaches improve upon isotropic diffusion by making the blurring a function of the image gradient. This selective smoothing preserves edges while reducing noise. Typically, \(D\) is a gradient-dependent diffusion coefficient, chosen as a monotonically decreasing function of \(\|\nabla u\|\), such as
\[
\begin{aligned}
D\left(\|\nabla u\|\right) &= \frac{1}{1+{\left(\|\nabla u\| / \kappa\right)}^2} \quad \text{or} \\
D\left(\|\nabla u\|\right) &= e^{-{\left(\|\nabla u\| / \kappa\right)}^2},
\end{aligned}
\]
where \(\kappa \in \mathbb{R}\) is a parameter that controls the sensitivity to edges. Anisotropic diffusion was first introduced in image processing in the seminal Perona-Malik equation, which is a nonlinear variant of the diffusion equation that selectively smooths regions with low gradients while preserving edges in high-gradient areas without significant loss of image details.

\subsection{Wave Equation}
This equation is used to describe the propagation of waves, such as sound, light, or water waves in certain media. The space-time propagation can be modeled as:  
\[
\frac{1}{c^2}\frac{\partial^2 u(x,y,t)}{\partial t^2} - \nabla^2 u(x,y,t) = R(x,y,t),
\]
where \(c\) is the wave speed and \(R(x,y,t)\) represents external forces or energy input, such as a speaker emitting sound waves.  

The form of the equation without the source term is more common:
\[
\frac{\partial^2 u}{\partial t^2} = c^2 \Delta u.
\]

This is a hyperbolic-type, second-order, linear with constant coefficients, and typically homogeneous PDE.

\subsection{Cahn-Hillard Equation}
This equation describes the dynamics of concentration fields undergoing phase separation and is commonly used for simulating materials or binary mixtures of two components subject to spinodal decomposition. Under various assumptions \cite{Wilczek2015}, it can be represented by the following PDE:  
\[
\frac{\partial c}{\partial t} = D \nabla^2 (-\nabla^2 c + c + c^3)= D \nabla^2 \mu(c).
\]
where \(D\in \mathbb{R}\) is a diffusion coefficient, \(c(x, y, t)\) is the concentration of one of the binary components, which is governed by a balance of diffusion and free energy, where the free energy depends on the local concentration and the interactions between phases, and \(\mu(c)\) is the chemical potential derived from the variation of a free energy functional.   

This equation is used to simulate the formation of microstructures and coarsening, where regions of uniform concentration merge to reduce the number of boundaries.  

In image processing, where \(c(x,y,t)=u(x,y,t)\) represents the intensity or value of each pixel, we think of the concentration of the substance as the image intensity. Therefore, the goal is to separate the substance’s distribution into low-texture and low-contrast regions to enhance the perceptual quality of the image. We will see in the experimental setting that this equation may be useful for image enhancement and sharpening, although it has already been used in the literature for image inpainting \cite{Gillette2006}, it has not been used in the particular form presented here.  

This is a parabolic-type, second-order, nonlinear PDE with constant or variable coefficients, and typically homogeneous.

\subsection{Burgers Equation}
The Burgers’ equation without viscosity (non-viscous) is a nonlinear hyperbolic PDE that models the evolution of shock waves in fluid dynamics:
\[
\frac{\partial u}{\partial t} + u_x \frac{\partial u}{\partial x} + u_y \frac{\partial u}{\partial v} = 0,
\]
where \( u(x,y,t) \) is the fluid velocity field and \( u_x \), \( u_y \) are the components of the velocity in the \( x \) and \( y \) directions, respectively. The equation describes the advection (transport) of \( u \) by its own gradient. It is used for modeling inviscid (non-viscous) flow and is often solved using finite-difference methods, such as the upwind gradient scheme, favoring stability. This is a hyperbolic-type, second-order, nonlinear, with constant coefficients, and typically homogeneous PDE.

When the viscosity term is included in Burgers’ equation, it takes the form:
\[
\frac{\partial u}{\partial t} + c_x \frac{\partial u}{\partial x} + c_y \frac{\partial u}{\partial v} = \nu \left( \frac{\partial^2 u}{\partial x^2} + \frac{\partial^2 u}{\partial y^2} \right),
\]
where \( \nu \) is the kinematic viscosity, which controls the diffusion of the fluid. This equation models the diffusion of a viscous fluid while accounting for advection and dissipation. This form of the equation is parabolic because it includes a diffusion term (the second derivative), and solutions will tend toward a smooth profile due to the diffusive term.

We have chosen to solve these hyperbolic PDEs using an upwind differences scheme, useful for advection-type equations, i.e., problems that involve the propagation of quantities or where there is a flow direction, such as information traversing along image gradients. The basic idea of this scheme is that it calculates the gradient in the direction opposite to the flow. The first-order upwind differences use:
\[
u_i^{n+1} = u_i^n - a \frac{\Delta t}{\Delta x} \left( u_i^n - u_{i-1}^n \right) \quad \text{for} \quad a > 0,
\]
\[
u_i^{n+1} = u_i^n - a \frac{\Delta t}{\Delta x} \left( u_{i+1}^n - u_i^n \right) \quad \text{for} \quad a < 0,
\]
where \( u_i^n \) denotes the quantity value at the \( i \)-th instant, \( \Delta t \) is the artificial time step, \( \Delta x \) is the spatial step (a pixel size \( \Delta x = \Delta y = 1 \) in 2D), and \( a \) is the velocity or flow speed, which represents how fast the quantity is transported in the system; its sign represents the flow direction.

The scheme requires fulfilling the Courant–Friedrichs–Lewy (CFL) condition (in 2D):
\[
c = \frac{a_x \Delta t}{\Delta x} + \frac{a_y \Delta t}{\Delta y} \le c_{\max},
\]
where \( c_{\max} = 1 \) for the explicit Euler time-marching method utilized.

\subsection{Transport Equation}
The transport (or advection) equation models the transport of a scalar quantity \( u \) along the direction of a velocity field. Its general form is:
\[
\frac{\partial u}{\partial t} + c_x \frac{\partial u}{\partial x} + c_y \frac{\partial u}{\partial v} = 0,
\]
where \( u(x,y,t) \) is the scalar quantity being transported and \( c_x \), \( c_y \) are the components of the velocity in the \( x \) and \( y \) directions, respectively. It is important to note that the form is similar to Burger’s non-viscous equation, but in this case the advection term is linear.

This is a hyperbolic-type, first-order, linear, with constant or variable coefficients, and typically homogeneous PDE.

\subsection{Liouville Equation}
The Liouville equation is commonly used in statistical mechanics and the dynamics of Hamiltonian systems. It plays a central role in statistical mechanics and Hamiltonian systems by describing how the probability density of particles evolves over time in phase space, ensuring the conservation of probability. In \( \mathbb{R}^3 \), phase space itself is a \( 6N \)-dimensional mathematical construct where the state of a system with \( N \) particles is represented by points defined by both position and momentum coordinates. This space is crucial for understanding the dynamics of the system, with the Liouville equation governing how the distribution of phase points moves through this space, analogous to the flow of a compressible fluid. In equilibrium, this results in a steady distribution, while in non-equilibrium systems, it dictates the evolution of particle interactions. The equation is fundamental in describing both equilibrium and nonequilibrium systems and is closely related to the canonical ensemble, where multiple systems are represented in a single phase space, all governed by the same underlying dynamics \cite{Mortimer}, and its general form is:
\[
\frac{\partial \rho(r, v, r)}{\partial t} + \nabla \cdot \left( \rho(r, v, r) \mathbf{v} \right) = 0,
\]
where \( \rho \) represents the probability density or distribution function—essentially the probability density of particles at position \( r \) with velocity \( v \) at time \( t \)—(in particle dynamics it represents the density of particles in phase space), and \( \mathbf{v} \) is the velocity vector of the particles, which describes the flow of the system. This equation governs how the distribution of particles evolves over time as a result of the dynamics of the system, or equivalently how probability density changes over time. It is first-order in time and ensures the conservation of probability in dynamic systems (no creation or destruction of probability in a closed system). A first-order stationary-state resolution approach gives: 
\[
\rho_{i,j}^{k+1} = \rho_{i,j}^{k} - \Delta t \cdot \left( v_x \frac{\partial \rho}{\partial x} + v_y \frac{\partial \rho}{\partial y} \right).
\]
This is an elliptic-type, second-order, nonlinear, with variable coefficients, and typically homogeneous PDE.

\subsection{Korteweg-de Vries Equation}
The Korteweg-de Vries (KdV) equation was first derived as a model for shallow water waves and later became central to the study of solitary waves, or solitons. Solitons are stable, localized waveforms that retain their shape and speed over time, a property that has been experimentally observed in various physical systems, including fluid dynamics and plasma physics. It has been used to model internal waves at the interface between two layers of fluids with different densities. Extensive experimental validation of its theory has been achieved, for example, in the study of wave profiles and particle velocities in oceanography, where the equation successfully predicts the formation and propagation of large-amplitude waves in regions such as the Andaman Sea \cite{Crighton1995}.

This equation describes solitary waves in dispersive media; considering dispersion in both directions, it can be formulated as:
\[
\frac{\partial u}{\partial t} + u\left( \frac{\partial u}{\partial x} + \frac{\partial u}{\partial y} \right) + \alpha \left( \frac{\partial^3 u}{\partial x^3} + \frac{\partial^3 u}{\partial y^3} \right) = 0,
\]
where \( u(x,y,t) \) is the dependent variable representing the wave height or profile of the solution (e.g., the wave height in 2D), the second term is the nonlinear advection term, representing dependence of \( u \) on its spatial derivative, the third term is the dispersion term, related to the curvature of the wave in both directions, and \( \alpha \in \mathbb{R} \) is a constant, typically \( \alpha = 6 \) by convention. We use an upwind differences scheme. The quantity \( u \) can represent various physical quantities, such as the displacement of the water surface from its equilibrium height. In image processing, it can correspond to the transfer of gradient information along the isophotes during an inpainting process or the redistribution of noise in piecewise smooth regions during denoising.

This is a hyperbolic-type, third-order, nonlinear, with constant coefficients, and typically homogeneous PDE.

\subsection{Kuramoto-Sivashinsky Equation}
This equation describes the nonlinear evolution of the Benjamin–Feir instability, which is a phenomenon in which nonlinearities amplify deviations from a periodic waveform, creating spectral sidebands and eventually causing the waveform to break up into a series of pulses. In addition, it has been studied as a model for interfacial instabilities, such as those in flame fronts and solidification processes. Intuitively, this equation describes spatio-temporal chaos and is often used in models of turbulent fluid dynamics. For a single scalar variable \( u(x,t) \), the equation writes:
\[
u_t + u_{xx} + u_{xxxx} + \frac{1}{2}(u_x)^2 = 0,
\]
and differentiating with respect to \(x\) and setting \( v = u_x \) leads to an alternative form analogous to the Navier-Stokes equations:
\[
v_t + v_{xx} + v_{xxxx} + v v_x = 0,
\]
where the last three terms correspond, respectively, to energy input at large scales, dissipation at small scales, and nonlinear advection \cite{Dawes2006}. The equation can be generalized to higher dimensions; one possibility in spatially periodic domains is:
\[
u_t + \Delta u + \Delta^2 u + \frac{1}{2} |\nabla u|^2 = 0,
\]
where \( \Delta^2 = \nabla^2 \nabla^2 \) is the biharmonic operator.

Specifically for an image \( u(x,y,t) \), the equations can be formulated as follows:
\[
\frac{\partial u}{\partial t} + u \left( \frac{\partial u}{\partial x} + \frac{\partial u}{\partial y} \right) + \left( \frac{\partial^2 u}{\partial x^2} + \frac{\partial^2 u}{\partial y^2} \right) + \left( \frac{\partial^4 u}{\partial x^4} + \frac{\partial^4 u}{\partial y^4} \right) = 0,
\]
and involves diffusive terms (both second- and fourth-order in space), as well as nonlinear advection terms in both the \(x\) and \(y\) directions. The high-order terms describe a strong dissipative (diffusive) process, whereas the second-order derivative term represents a standard weaker diffusion process, and the first-order ones describe the nonlinear interactions of the wave in both spatial directions. We use an upwind differences scheme. 

The independence of explicit parameters in the Kuramoto-Sivashinsky (K-S) equation is highly relevant for image processing, as it allows for flexible, automatic adaptation to diverse image conditions, reducing the need for manual adjustments and enhancing efficiency in tasks like deblurring and other restoration tasks.

This is a parabolic-type, fourth-order, nonlinear, with constant coefficients, and typically homogeneous PDE.

\subsection{Fluid Dynamics Equations}
These are also nonlinear equations in physics heavily reliant on numerical solutions under most conditions, as the sources and their quantities are related to the solutions:
\[
\begin{matrix}
   \frac{\partial \mathbf{v}}{\partial t} + \mathbf{v} \cdot \nabla \mathbf{v} + \frac{1}{\rho} \nabla P - \eta \nabla^2 \mathbf{v} = 0,  \\
   \frac{\partial \rho}{\partial t} + \nabla \cdot \rho \mathbf{v} = 0,  \\
   f(P, \rho) = 0.  \\
\end{matrix}
\]
Here the first equation is the Navier–Stokes equation, in which \( \mathbf{v} \) is the velocity, \( \rho \) the density, \( P \) the pressure, and \( \eta \) the kinetic viscosity of the fluid. The Navier–Stokes equation can be derived from Newton's equation for a small element in the fluid. The second equation is the continuity equation, which is the result of mass conservation. The third is the equation of state, which can also involve temperature as an additional variable in many cases \cite{Pang}.

Navier-Stokes equations have been applied in \cite{Bertalmío} to image inpainting to restore missing or corrupted parts of images by estimating surrounding pixel values, often involving solving nonlinear equations. This approach treats image intensity as a stream function for two-dimensional fluid flow, propagating information from surrounding areas to missing regions. The Laplacian of image intensity acts as vorticity, and the gradient vectors are matched at the boundaries. Additionally, these equations have been used for superresolution, enhancing the resolution of magnified images. Using computational fluid dynamics (CFD), this method introduces well-established numerical techniques to improve image inpainting and resolution, with potential for further improvements through alternative algorithms and dynamic fluid instabilities.

\subsection{Maxwell’s Equations}
The Maxwell equations describe the fundamental behavior of electric and magnetic fields in space and time. In their most general form, Heaviside’s differential formulation, they are a set of four coupled partial differential equations that govern electromagnetism:
\[
\begin{matrix}
   \nabla \cdot \mathbf{E}=\frac{\rho }{{{\varepsilon }_{0}}}  \\
   \nabla \cdot \mathbf{H}=0  \\
   \nabla \times \mathbf{E}=-\frac{\partial \mathbf{H}}{\partial t}  \\
   \nabla \times \mathbf{H}={{\mu }_{0}}\mathbf{J}+{{\mu }_{0}}{{\varepsilon }_{0}}\frac{\partial \mathbf{E}}{\partial t}.  \\
\end{matrix}
\]
Maxwell-Heaviside’s (M-H) equations, fundamental to electromagnetism, can be categorized into elliptic and hyperbolic equations, each governing different aspects of electromagnetic phenomena. The first, Gauss's law for electric fields (elliptic), states that electric fields are generated by electric charges, the field diverging from positive charges and converging at negative ones. It highlights how the electric field is related to the presence of charge distributions, laying the foundation for understanding electrostatic fields. The second, Gauss's law for magnetism (elliptic), asserts that magnetic fields have no monopoles and always form closed loops. This indicates that magnetic flux is conserved in nature, defining the behavior of magnetic fields, which are generated by moving charges or magnetic dipoles. The third, Faraday's law of induction (hyperbolic), establishes that a changing magnetic field generates a circulating electric field. This principle is essential to the operation of electric generators and transformers and shows how a time-varying magnetic flux induces an electromotive force (EMF) in a closed loop of wire, which is central to the generation of electrical energy. The fourth, Ampère-Maxwell Law (hyperbolic), describes how electric currents and time-varying electric fields produce magnetic fields. It is key to understanding electromagnetism and the propagation of electromagnetic waves, with the term displacement current introduced by Maxwell, modifying Ampère's law to account for time-varying electric fields. 

Together, these equations predict the existence of electromagnetic waves, oscillating electric and magnetic fields that propagate through space at the speed of light. Under certain considerations, the propagation of these waves can be derived from Maxwell’s equations, revealing that electric and magnetic fields are intrinsically linked and move together as waves. This concept not only explains electromagnetic radiation, such as light, radio waves, X-rays, and microwaves, but also underscores that such waves do not require a medium like sound waves do, as was once thought with the “ether” theory. Maxwell’s equations therefore provide a comprehensive framework to understand the behavior of electromagnetic fields and waves, bridging the gap between electricity, magnetism and optics.

The intensity of an electromagnetic wave can be calculated as the average of the amount of energy that flows per unit of time through a unit area perpendicular to the energy flow; thus the energy is related to the Umov-Poynting vector
\(\mathbf{S} = (\mathbf{E} \times \mathbf{H}) / {\mu}\), which represents the directional energy flux or power flow of an electromagnetic field, \(\mu\) being the medium permeability. The total intensity \cite{Cao2021}. By incorporating this intensity as a gradient-based advection term, it becomes possible to guide an image restoration process by emphasizing energy flow along significant gradients, preserving critical structures while suppressing noise.

Hence, we synthesize the four equations into a compact, simple but flexible, and heuristic formulation that incorporates all basic needs for solving various image processing tasks in a PDE-based approach. Specifically, this implementation solves a modified Maxwell-Heaviside PDE, where the gradient of the image is treated as the electric field \( \mathbf{E} \), and the curl of the gradient is treated as the magnetic field \( \mathbf{H} \). This model integrates diffusion, advection, and reaction processes. The main PDE used here is of the form:
\[
\frac{\partial u}{\partial t} = \alpha \nabla^2 u - \beta \nabla \cdot (\mathbf{E} \times \mathbf{H}) + \gamma |\mathbf{E}|^2,
\]
where \( u \) is the image or scalar field at time \( t \), \( \alpha \) is the diffusion coefficient (control over smoothing), \( \beta \) is the advection coefficient (controls the transport or interaction between the electric and magnetic fields), \( \gamma \) is the reaction coefficient (related to sharpness enhancement), \( \mathbf{E} = (E_1, E_2) \) is the electric field (approximated by the gradient of the image), \( \mathbf{H} = (H_1, H_2) \) is the magnetic field (computed also from the image gradients), \( \nabla^2 \) acts as a diffusion operator, \( |\mathbf{E}| \) acts as a measure of the total electromagnetic field (EM), and \( \nabla \cdot (\mathbf{E} \times \mathbf{H}) \) is the divergence of the cross product of \( \mathbf{E} \) and \( \mathbf{H} \) acting as an advection-like term, defined as:
\[
\nabla \cdot (\mathbf{E} \times \mathbf{H}) = \frac{\partial (E_x H_y - E_y H_x)}{\partial x} + \frac{\partial (E_x H_y - E_y H_x)}{\partial y}.
\]

The electric field \( \mathbf{E} = (E_1, E_2) \) is approximated by the gradient of the image, i.e.,
\[
\mathbf{E} = \left( \frac{\partial u}{\partial x}, \frac{\partial u}{\partial y} \right),
\]
approximating derivatives using convolutions with central difference kernels
\[
k_x = \frac{1}{2} \begin{bmatrix} -1 & 0 & 1 \end{bmatrix}
\quad \text{and} \quad
k_y = \frac{1}{2} \begin{bmatrix} -1 & 0 & 1 \end{bmatrix}^T.
\]
On the other hand, the magnetic field \( \mathbf{H} = (H_1, H_2) = (H, H) \) is computed from the gradients of the electric field as
\[
\mathbf{H} = \left( \frac{\partial E_y}{\partial x} - \frac{\partial E_x}{\partial y}, \frac{\partial E_y}{\partial x} - \frac{\partial E_x}{\partial y} \right).
\]
The curl of the gradient being zero simplifies the mathematical treatment of the PDE, compensating for numerical imprecisions and ensuring a more stable solution. This behavior mirrors the orthogonality of electric and magnetic fields in electromagnetism, allowing the image gradient to behave naturally within the system. By these approximations, the proposed approach leads to a consistent and physically accurate solution that aligns with the expected behavior of electromagnetic fields. Moreover, in the implementation, we incorporate a parameter which controls whether the fields are treated as fixed (initial condition) or recalculated at each iteration during the PDE evolution.

The \( \left| \mathbf{E} \right| \) term can be calculated in two different ways depending on the particular task:
\begin{itemize}[label=-]
    \item By means of the Vector Magnitude Equation:
    \[
    EM = \sqrt{E_x^2 + E_y^2 + 2H^2}
    \]
    \item By means of the Energy Density Method:
    \[
    EM = \frac{1}{2} \left( \varepsilon_0 \left\| \mathbf{E} \right\|^2 + \mu_0 H^2 \right),
    \]
    where \( \varepsilon_0 \) is the permittivity of free space, \( \mu_0 \) is the permeability of free space, and \( \left\| \mathbf{E} \right\|^2 = E_x^2 + E_y^2 \) denotes the squared \( l^2 \)-norm.
\end{itemize}

Then, in order to solve this PDE, we choose to follow two alternative approaches:
\begin{itemize}[label=-]
    \item Using the Euler explicit method:
    \[
    u^{k+1} = u^k + \Delta t \cdot \left( \alpha \nabla^2 u - \beta \nabla \cdot (\mathbf{E} \times \mathbf{H}) + \gamma |\mathbf{E}|^2 \right).
    \]
    \item Using the Runge-Kutta fourth-order method (RK4) for time integration, involving four steps per iteration:
    \[
    \begin{matrix}
       k_1 = \alpha \nabla^2 u - \beta \nabla \cdot (\mathbf{E} \times \mathbf{H}) + \gamma |\mathbf{E}|^2,  \\
       k_2 = \alpha \nabla^2 \left( u + 0.5 k_1 \Delta t \right) - \beta \nabla \cdot (\mathbf{E} \times \mathbf{H}) + \gamma |\mathbf{E}|^2,  \\
       k_3 = \alpha \nabla^2 \left( u + 0.5 k_2 \Delta t \right) - \beta \nabla \cdot (\mathbf{E} \times \mathbf{H}) + \gamma |\mathbf{E}|^2,  \\
       k_4 = \alpha \nabla^2 \left( u + k_3 \Delta t \right) - \beta \nabla \cdot (\mathbf{E} \times \mathbf{H}) + \gamma |\mathbf{E}|^2,  \\
    \end{matrix}
    \]
    the updated image being as follows:
    \[
    u^{k+1} = u^k + \frac{\Delta t}{6} \left( k_1 + 2k_2 + 2k_3 + k_4 \right),
    \]
    where \( \Delta t \in \mathbb{R} \) is an artificial time-step.
\end{itemize}

When \( \alpha \ne 0, \beta = \gamma = 0 \), the equation models classical heat diffusion, where the field spreads out over time. This is ideal for removing noise or performing image inpainting, as the diffusion term smooths the field and eliminates irregularities, leading to a gradual homogenization of the field values. 

When \( \beta \ne 0, \alpha = \gamma = 0 \), the equation describes advection or a pure transport process. The electromagnetic field is carried or transported through space due to the interaction between \( \mathbf{E} \) and \( \mathbf{H} \). This term balances the effect of diffusion and reaction, and the equation models the movement of the field with minimal smoothing or sharpening, which is indicated to simulate the motion of fields or particles within a fluid or medium.

Finally, when \( \gamma \ne 0, \alpha = \beta = 0 \), the equation focuses on a pure reaction based on the magnitude of the total electromagnetic field. This is ideal for enhancing sharpness and clarity in the field, as the reaction term emphasizes regions with high electromagnetic intensity. This approach works well when the goal is to highlight features or edges within the field, essentially sharpening the image.

In combination, this PDE incorporates diffusion, advection, and reaction, where diffusion smooths noise, advection moves the field, and reaction sharpens key features. The relative magnitudes of the three parameters determine the balance between these behaviors and shape the evolution of the electromagnetic field in the system.

\section{Experiments, Results and Analysis}
\subsection{Image denoising}
\begin{figure*}[p!]
    \centering
    \includegraphics[width=\linewidth]{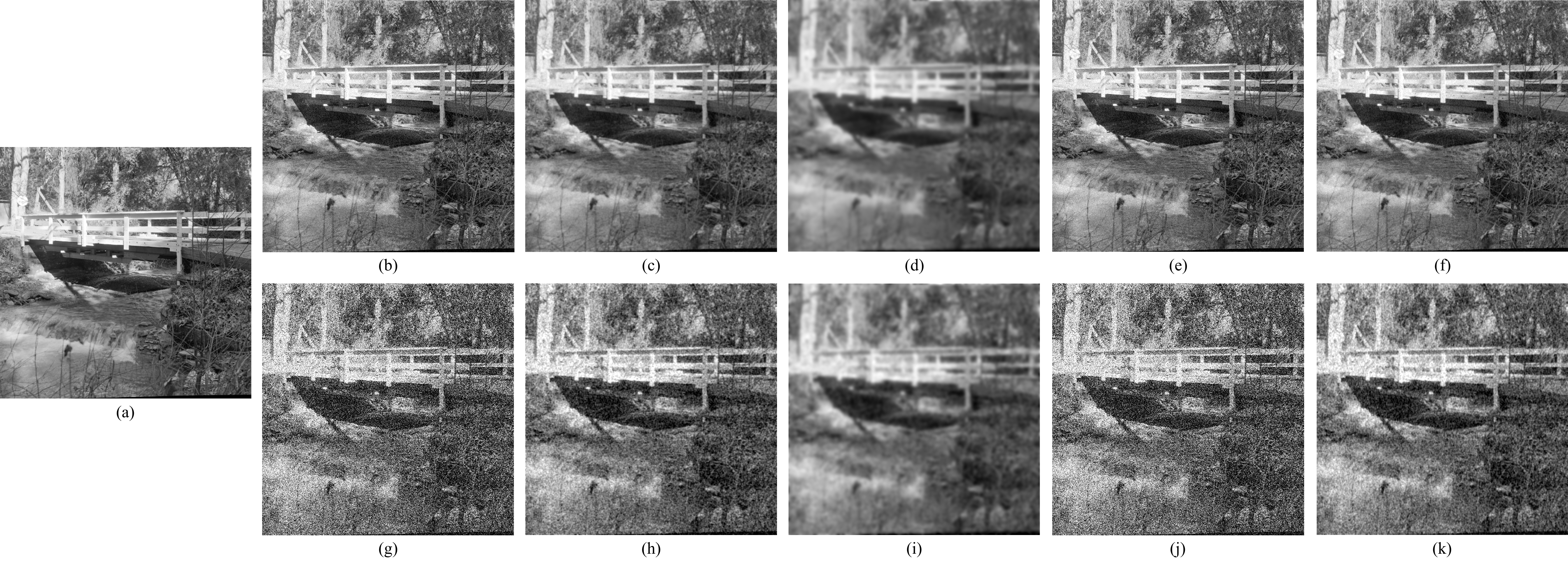}
    \caption{Denoising result with (a) bridge image. For   in the upper row, for   in the bottom row. (b, f) Heat, (c, g) Laplace, (d, h) P-M, (e, i) M-H.}
    \label{fig:Figure1}
\end{figure*}

\begin{figure*}[p!]
    \centering
    \includegraphics[width=\linewidth]{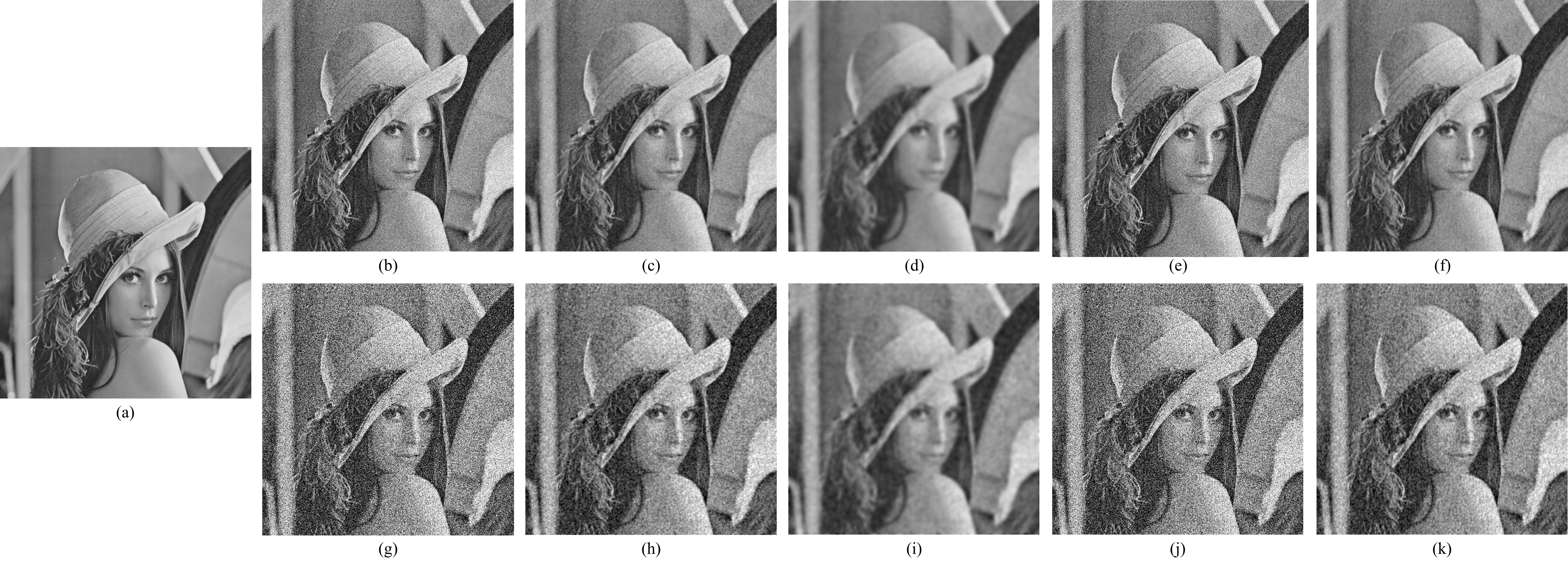}
    \caption{Denoising result with (a) Lena image. For   in the upper row, for   in the bottom row. (b, f) Heat, (c, g) Laplace, (d, h) P-M, (e, i) M-H.}
    \label{fig:Figure2}
\end{figure*}

To assess the effectiveness of the proposed denoising models in this study, we compared their performance against other widely used methods. The evaluation of these methods is based on three criteria: mean square error (MSE), peak signal-to-noise ratio (PSNR) and structural similarity index measure (SSIM). For details of these metrics, please refer to \cite{Yang2021}\cite{Zhao2024}\cite{Murata2022}.
White noise was added by introducing random noise (\textit{i.i.d.}) following a normal distribution with a mean of zero and a standard deviation of 0.1 and 0.25 (see Fig. \ref{fig:Figure1} and \ref{fig:Figure2}).
\begin{table}[h!]
    \centering
    \caption{Evaluation of the proposed PDEs and comparison with other denoising methods.}
    \includegraphics[width=\linewidth]{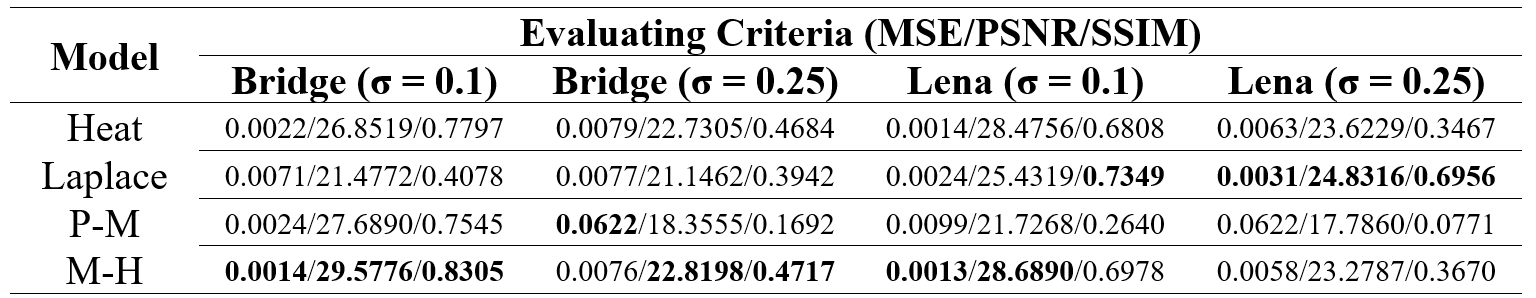}
    \label{fig:Table1}
\end{table}
\vspace{-5mm}

\subsection{Image deblurring}
In image deblurring/deconvolution, the goal is to restore a sharp and clear image from a blurred and (possibly) noisy one. To assess the effectiveness of the proposed reconstruction deblurring/deconvolution models in this study, we compared their performance against other widely used methods. The evaluation of these methods is based on two criteria: quality index (Q) and peak signal-to-noise ratio (PSNR). For details about these metrics please refer to \cite{Yang2021}\cite{Zhao2024}\cite{Murata2022}\cite{Zhou}.

A Gaussian kernel of size \( 11 \times 11 \) and a standard deviation \( \sigma = 3 \) were used in the blurring process of the images. In addition, we added white noise with a standard deviation of 0.05 to impose realism in the evaluation. The values of the image pixels are assumed to be within the range \([0, 1]\), i.e., 
\[
u: \Omega \subset \mathbb{R}^2 \to \mathbb{R}, \quad u(x, y) \in [0, 1].
\]
Also, we use periodic boundary conditions with a 3-pixel extension to avoid artifact effects at the edges of the images and later crop them to the original size after the restoration process.

\begin{figure*}[p!]
    \centering
    \includegraphics[width=\linewidth]{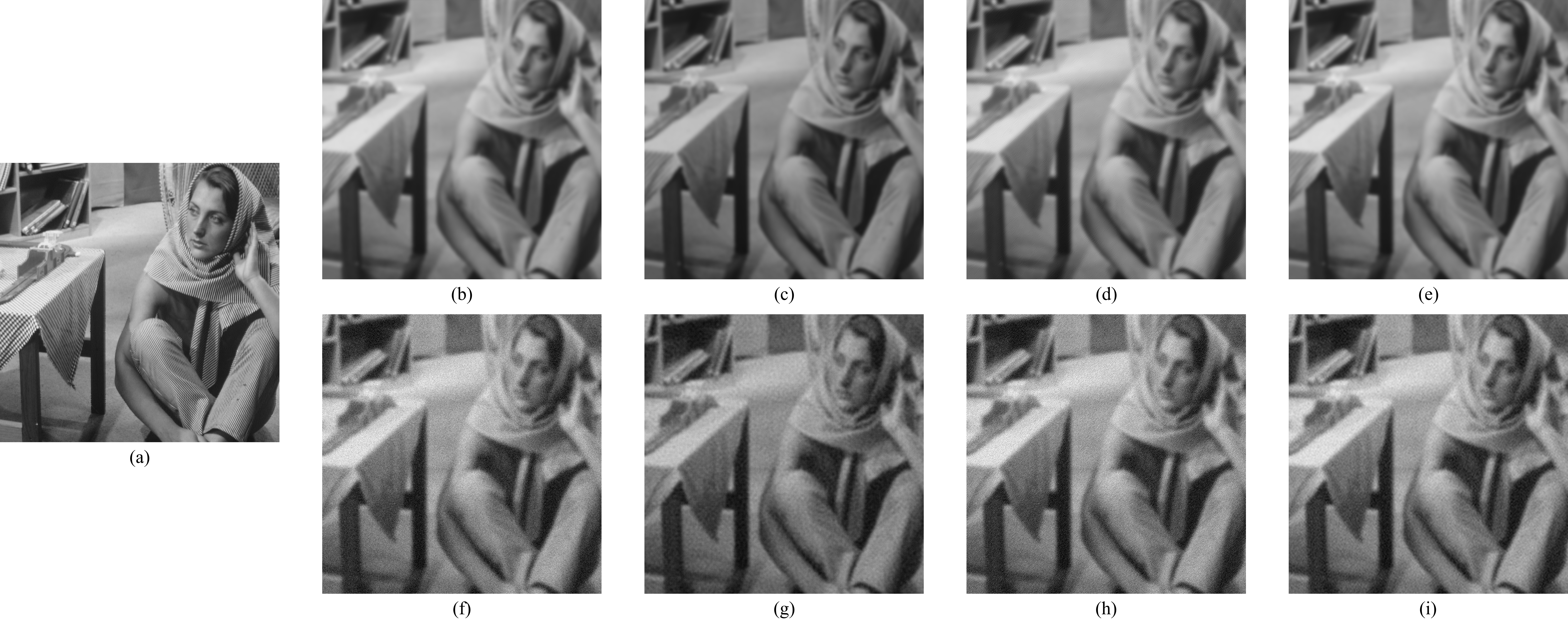}
    \caption{Deblurring result with (a) Barbara image. For blurry image in the upper row, for blurry and noisy image in the bottom row. (b, f) Input, (c, g) Tikhonov, (d, h) TV, (e, i) K-S.}
    \label{fig:Figure3}
\end{figure*}

\begin{figure*}[p!]
    \centering
    \includegraphics[width=\linewidth]{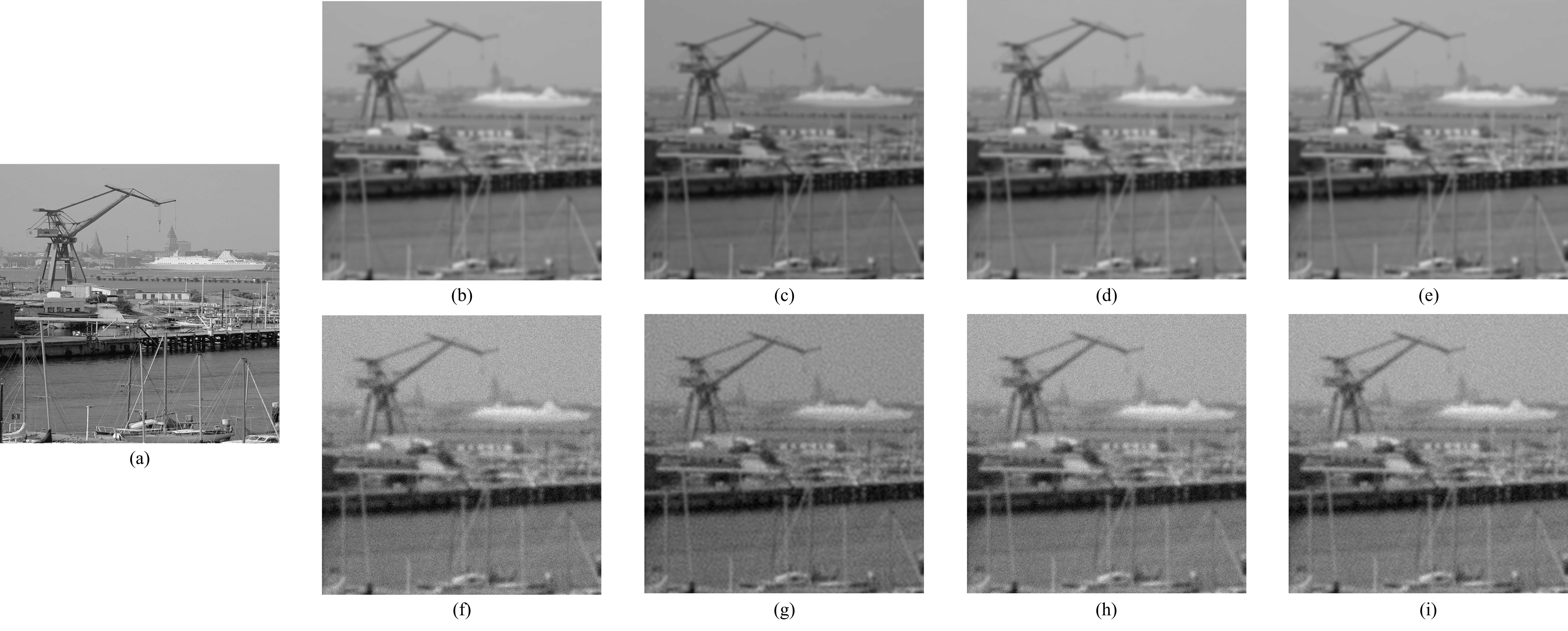}
    \caption{Deblurring result with (a) Kiel image. For blurry image in the upper row, for blurry and noisy image in the bottom row. (b, f) Input, (c, g) Tikhonov, (d, h) TV, (e, i) K-S.}
    \label{fig:Figure4}
\end{figure*}

We evaluated the proposed Kuramoto-Sivashinsky PDE against two classic regularization-based models, the Tikhonov and Total Variation (TV) variational methods; see Fig. \ref{fig:Figure3} and Fig. \ref{fig:Figure4}.
\begin{table}[h!]
    \centering
    \caption{Evaluation of the proposed PDE and comparison with other deblurring methods.}
    \includegraphics[width=\linewidth]{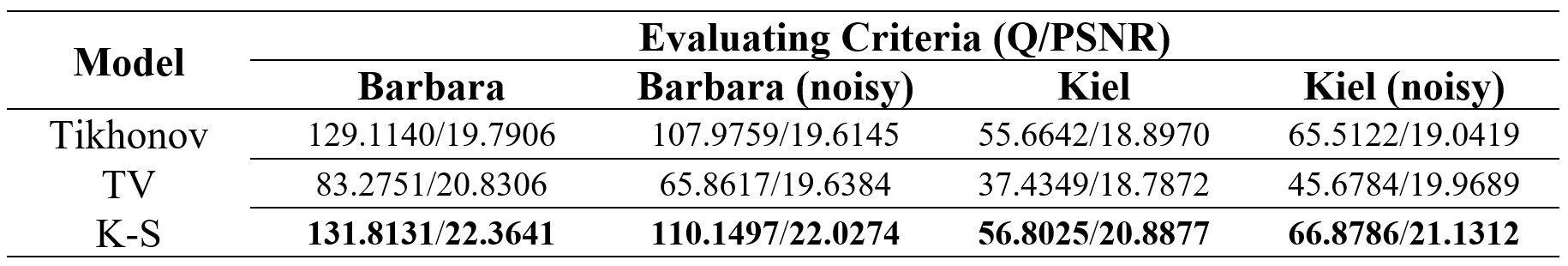}
    \label{fig:Table2}
\end{table}
\vspace{-5mm}

Unlike variational methods with regularization, such as TV and Tikhonov, which seem to be more prone to introducing high-frequency artifacts and often produces the staircase effect, the PDE-based method of Kuramoto-Sivashinsky tends to preserve piecewise smooth regions more effectively while also maintaining edges, leading to a more natural appearance in the restored image. Although the performance of the Tikhonov method and the Kuramoto-Sivashinsky method yield similar results in an objective analysis, in a subjective one, the visual aspect returned by the PDE is more pleasing in the sense that it presents better contrast and regions of intensity that are more consistent.

Moreover, a clear advantage of PDE-based methods is that they do not require an a priori estimate of the PSF that degraded the image for deconvolution; the proposed PDE allows tackling a broader spectrum of deblurring and deconvolution problems.

\subsection{Image enhancement}
One of the objectives of image enhancement, closely related to that of sharpening, is to modify attributes of an image, such as increasing contrast, extending the dynamic range, uniformizing the histogram or sharpening the image, to make it more suitable for a given task and a specific observer. 

\begin{figure*}[p!]
    \centering
    \includegraphics[width=\linewidth]{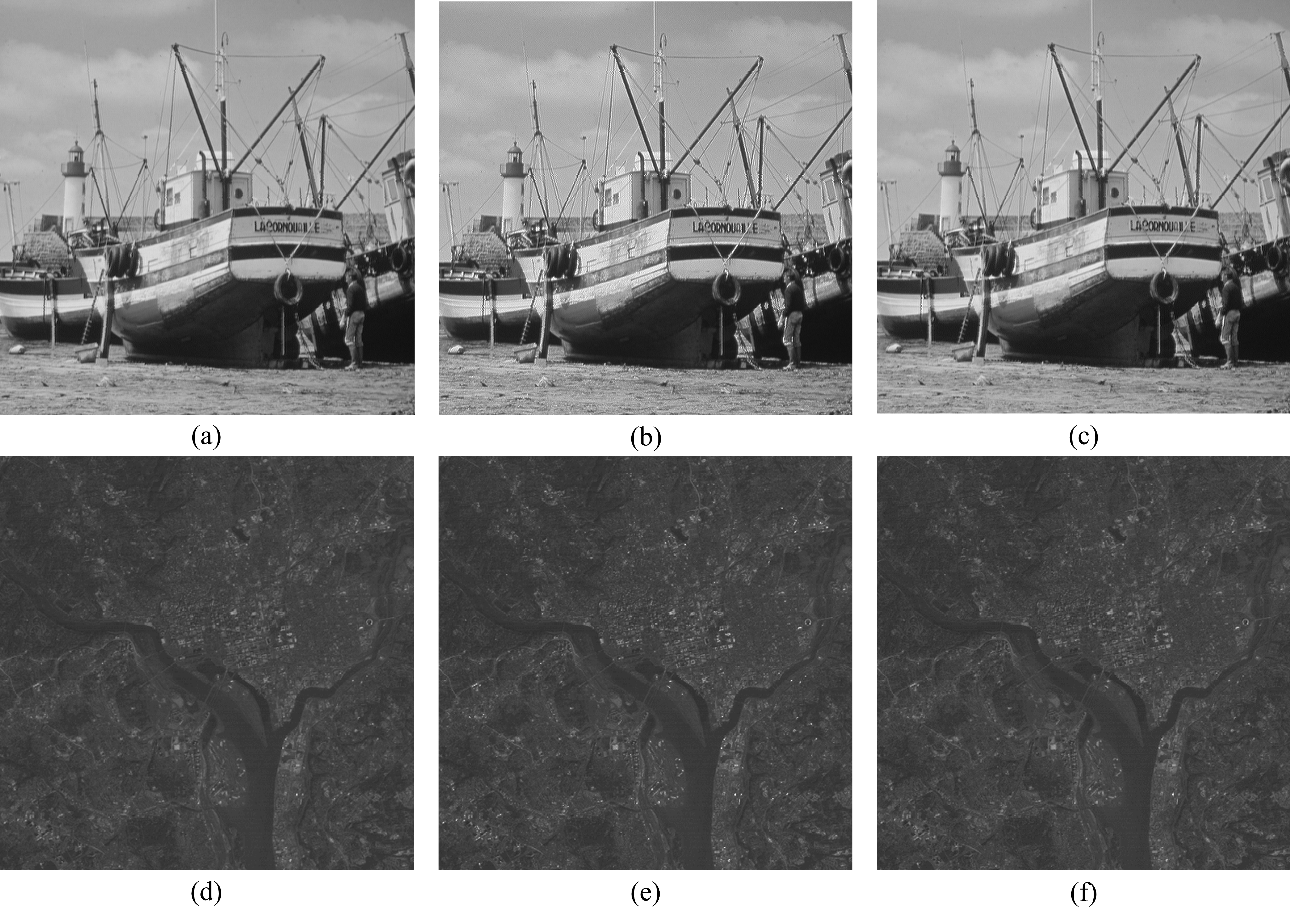}
    \caption{Enhancement results with (a, b) monochrome images. (b, e) C-H, (c, f) M-H.}
    \label{fig:Figure5}
\end{figure*}

\begin{figure*}[p!]
    \centering
    \includegraphics[width=\linewidth]{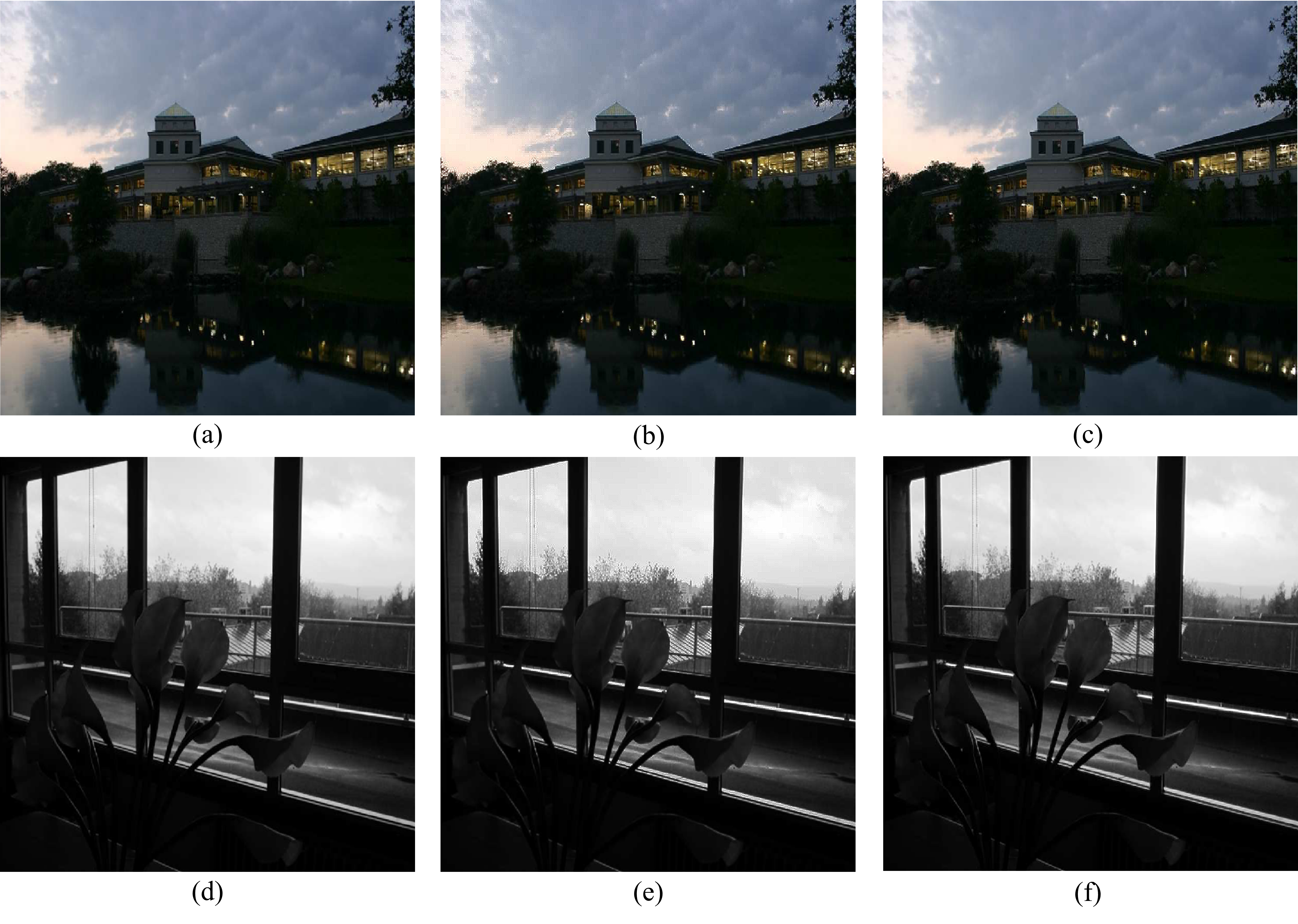}
    \caption{Enhancement results with (a, b) vector-valued images. (b, e) C-H, (c, f) M-H.}
    \label{fig:Figure6}
\end{figure*}

Among the PDEs we have proposed, the Cahn-Hilliard (C-H) and Maxwell-Heaviside equations are ideal, as it is possible to invert the behavior of the diffusion term in them and cause a “reverse diffusion” that sharpens the edges. The testing results (see Fig. \ref{fig:Figure5} and see Fig. \ref{fig:Figure6}) show that the evolution they trigger enhances fine gradients and mid-contrast areas, improving the quality or visual perception of the image without altering its mean grayscale level.

To evaluate the capabilities of the methods, we use the QV value metric proposed in \cite{Nasonov2011}, which is indicated for quality measurement of image enhancement algorithms. We also compute the PSNR, to ensure that the enhanced image remains sufficiently realistic compared to the original. 
\begin{table}[h!]
    \centering
    \caption{Evaluation of the proposed PDE for enhancing.}
    \includegraphics[width=\linewidth]{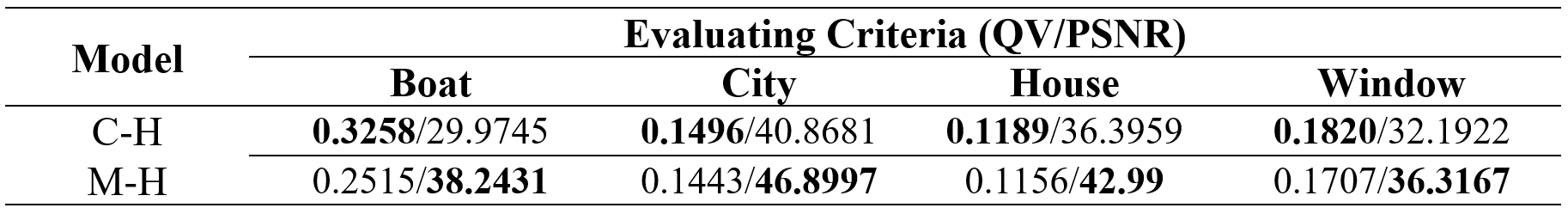}
    \label{fig:Table3}
\end{table}
\vspace{-5mm}

Although the viscous Burgers PDE seems to result in a better quality coefficient on average, which may indicate a greater richness in important gradients in the image, the Maxwell-Heaviside equation is much more faithful to the original image, which may be preferable in certain situations.

\subsection{Image inpainting}
\begin{figure*}[p!]
    \centering
    \includegraphics[width=\linewidth]{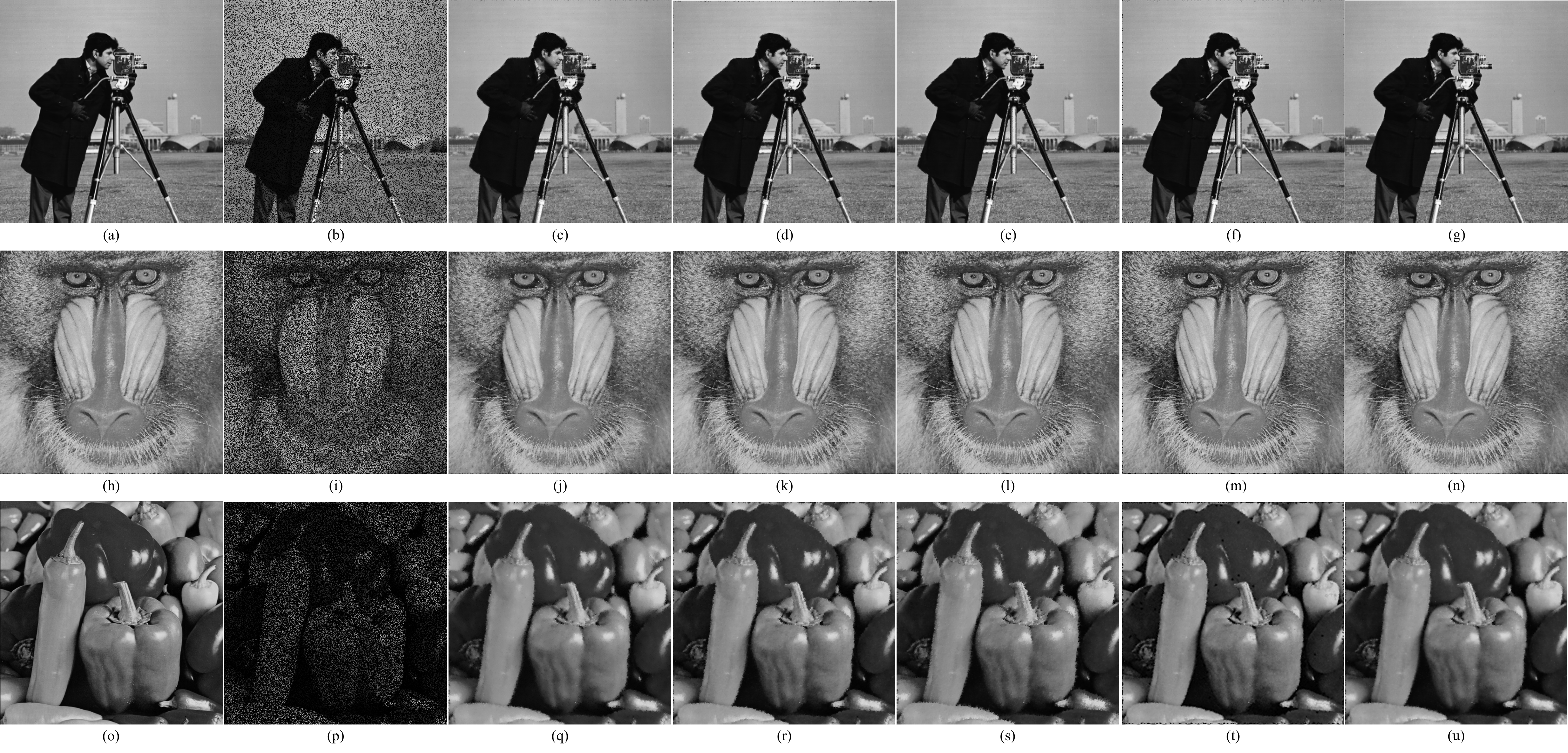}
    \caption{Inpainting results with (a, h, o) monchrome images. (b, i, p) damaged inputs, (c, j, q) Telea, (d, k, r) N-S, (e, l , s) BV, (f, m, t) C-H, (g, n, u) M-H.}
    \label{fig:Figure7}
\end{figure*}

In image inpainting, damaged parts or lost subdomains of the image are digitally reconstructed while maintaining their original look (see Fig. \ref{fig:Figure7}).

We compare two state-of-the-art techniques, Navier-Stokes and Telea’s inpainting, against three of our proposed physical-based PDE models, BV, C-H and M-H.
\begin{table}[h!]
    \centering
    \caption{Evaluation of the proposed PDEs and comparison with other inpainting methods.}
    \includegraphics[width=\linewidth]{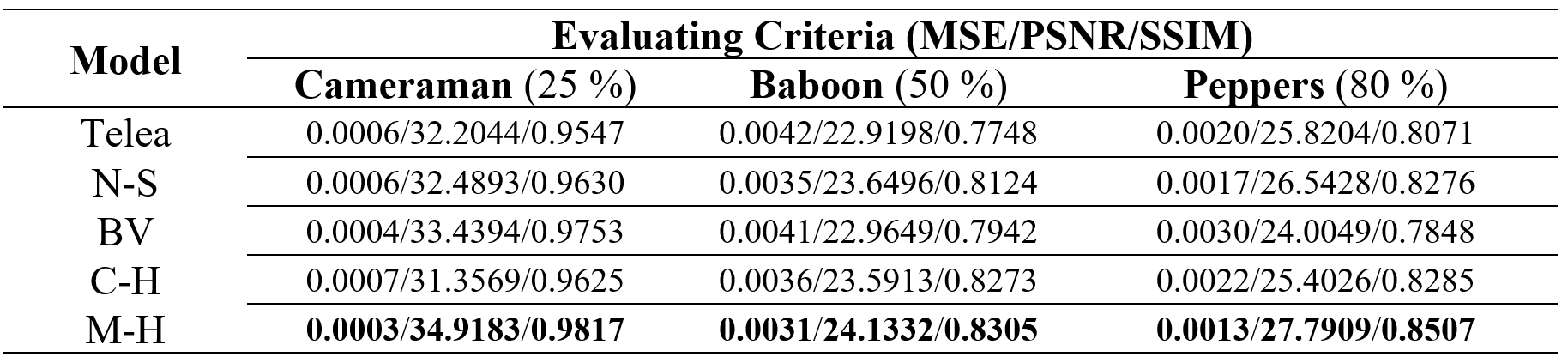}
    \label{fig:Table4}
\end{table}
\vspace{-5mm}

We observe that the \(\gamma\) parameter in the Cahn-Hilliard equation, which is related to the chemical potential, plays a critical role in improving inpainting in dark areas and regions with very low contrast. However, increasing this parameter excessively leads to the appearance of texture artifacts, overexposing the intensity of neighborhoods with high-frequency textures. 

It can be stated that the proposed PDE-based solutions, Cahn-Hilliard and particularly Maxwell-Heaviside, significantly outperform Navier-Stokes and Telea’s inpainting, in terms of restoration through the metrics considered. 

The Maxwell PDE is more effective than other inpainting methods in preserving fine gradients and high-frequency details. Additionally, it introduces a sharpening effect that enhances the quality of the reconstruction, making it more closely resemble the image before corruption. It also performs better at handling edges compared to the Navier-Stokes equation, although the latter is also based on a physical PDE.

\subsection{Edge detection}
\begin{figure*}[p!]
    \centering
    \includegraphics[width=\linewidth]{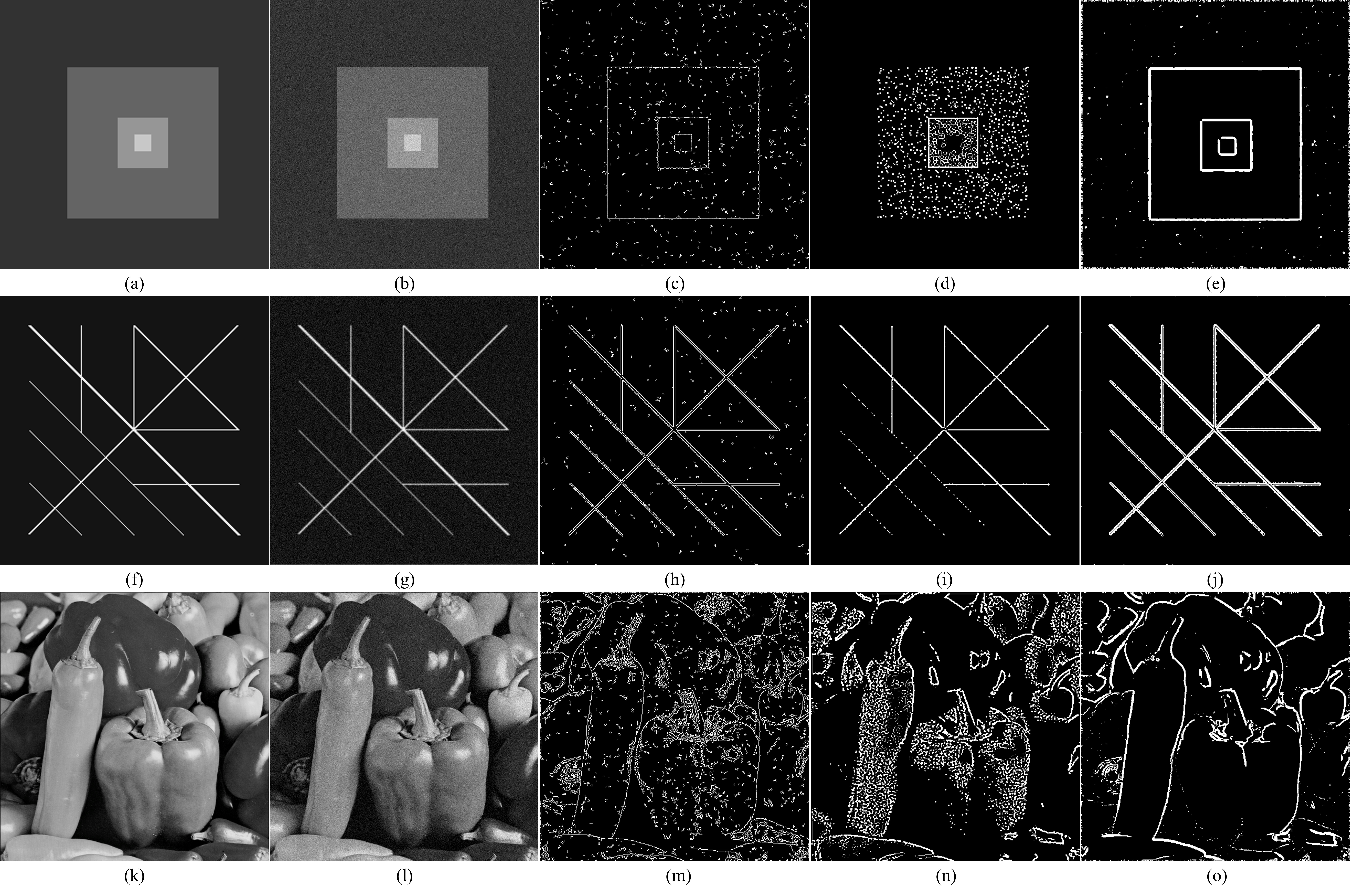}
    \caption{Results of edge detection with KdV and comparison with other methods. (a, f, k) Test images, (b, g, l) noisy images, (c, h, m) Canny, (d, I, n) LoG, (e, j, o) KdV.}
    \label{fig:Figure8}
\end{figure*}

The filtering using the KdV PDE results in a natural appearance of the image. This can be explained by the specific terms in the governing equations that contribute to this effect. For this reason, it may be more suitable than traditional techniques for edge detection such as Canny or LoG-based edge detection methods, as illustrated in Fig. \ref{fig:Figure8}.

\begin{figure*}[p!]
    \centering
    \includegraphics[width=\linewidth]{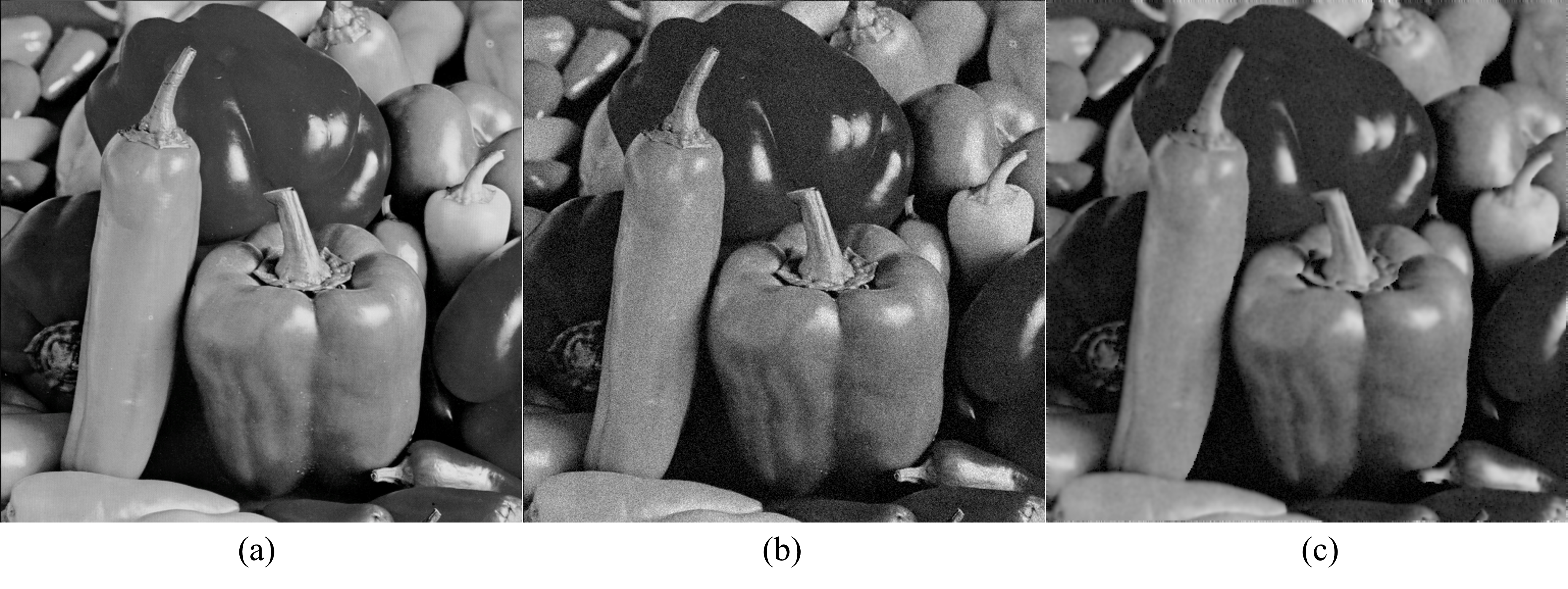}
    \caption{Detailing the filtering behavior offered by KdV. (a) Input, (b) noisy, (c) filtered.}
    \label{fig:Figure9}
\end{figure*}

The KdV PDE exhibits remarkable robustness to noise (as can be seen in Fig. \ref{fig:Figure9}), which proves advantageous for edge detection in images.

It is observed that, compared to classical methods like Canny and LoG, which often struggle with noisy images, the proposed PDE tends to enhance the strong edges of objects. This makes it particularly useful for applications such as segmentation or similar tasks, as it provides greater resilience to variable environmental conditions, due to its dispersion and advection terms.

\subsection{Other tasks}
Other tasks in image processing would include tone management, which adjusts the brightness and contrast of an image to enhance its visual appeal, color transfer, which matches the color tones of one image to another for stylistic or consistency purposes, cartooning, which simplifies images into stylized, cartoon-like versions and even visual effect transformations used to implement physical dynamic processes in images, without the need for a complex rendering engine. In addition, optimal transport is particularly useful in registration, enabling us to align multiple images into a common framework for accurate comparison or analysis. In the following, we present some promising results that could contribute to addressing numerous image processing challenges using the physical-based PDEs proposed in this work.

\begin{figure*}[p!]
    \centering
    \includegraphics[width=\linewidth]{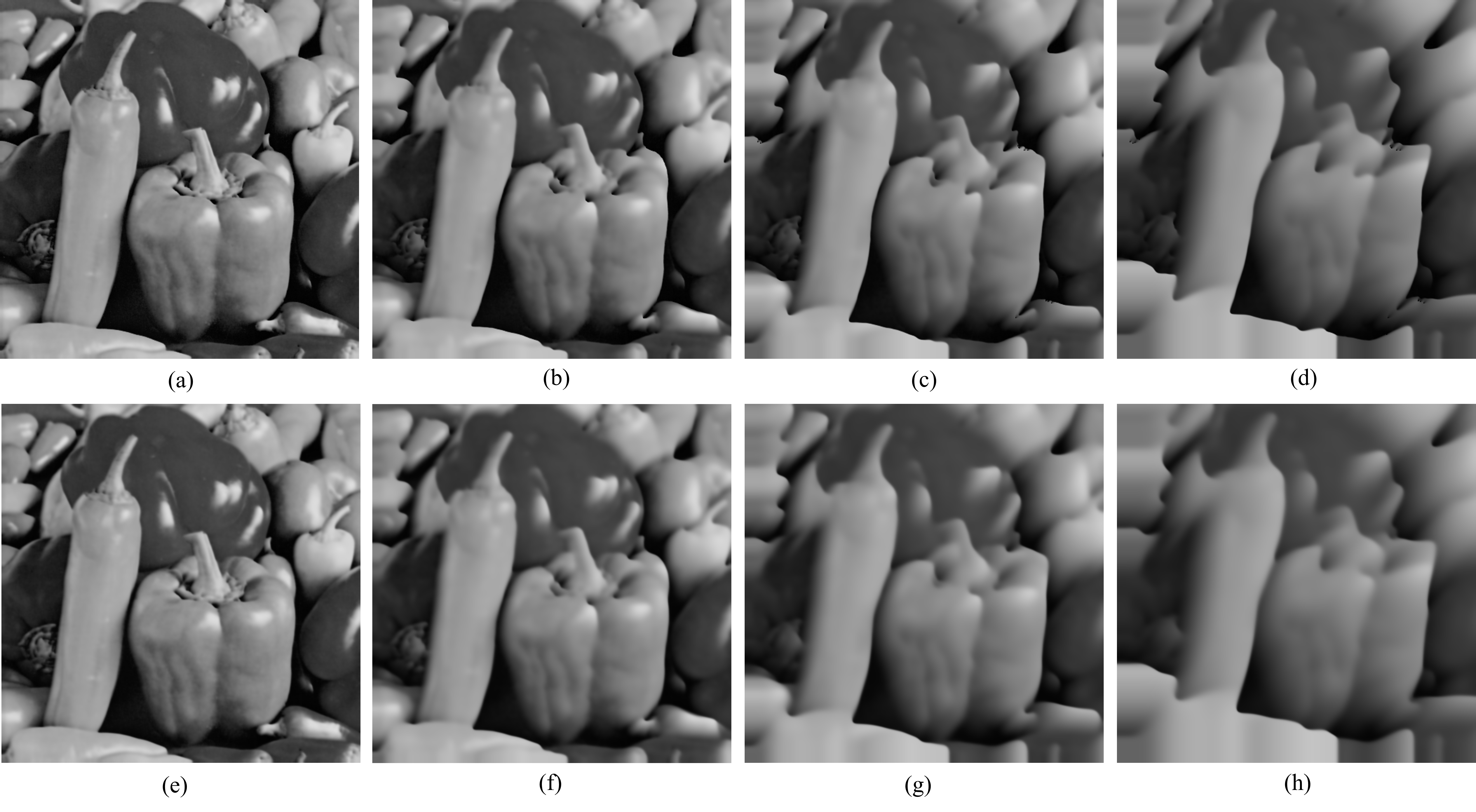}
    \caption{Behavior of Burger’s viscous (e-h) and non-viscous (a-d) PDEs.}
    \label{fig:Figure10}
\end{figure*}

Burgers’ viscous and non-viscous equations can be used for the decomposition of the image into cartoon/texture components, or it can also be extended for segmentation if a method is added to the PDE to control the directions in which the object contours are transported, guided, and restricted by the gradients in their neighborhood (see Fig. \ref{fig:Figure10} for a detailed view at different timestamps); it can also perform well in inpainting and magnification due to its advective qualities.

\begin{figure*}[p!]
    \centering
    \includegraphics[width=\linewidth]{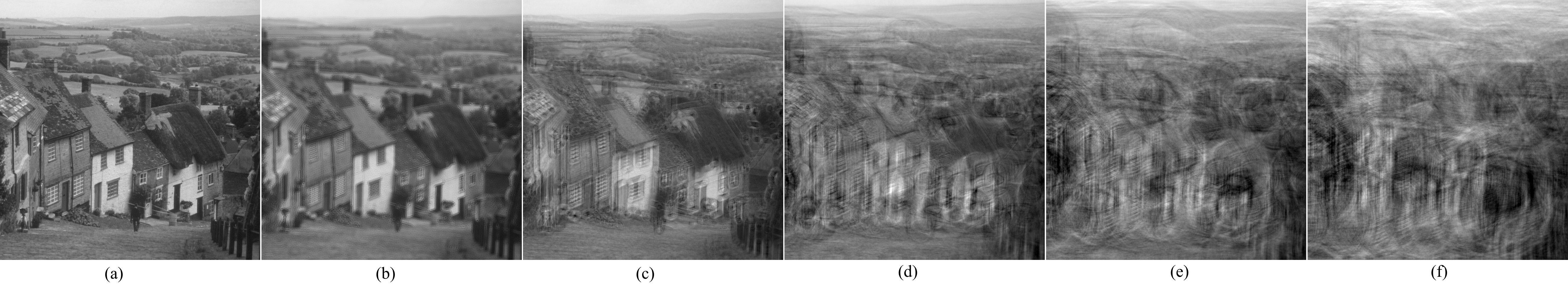}
    \caption{(a-f) Evolution of Goldhill image under of wave equation.}
    \label{fig:Figure11}
\end{figure*}

In Fig. \ref{fig:Figure11}, one can observe the evolution of the Goldhill image driven by the wave equation. In addition to its practical applications, it can also be used to create visual or artistic effects based on subjective interpretation.

\begin{figure*}[p!]
    \centering
    \includegraphics[width=\linewidth]{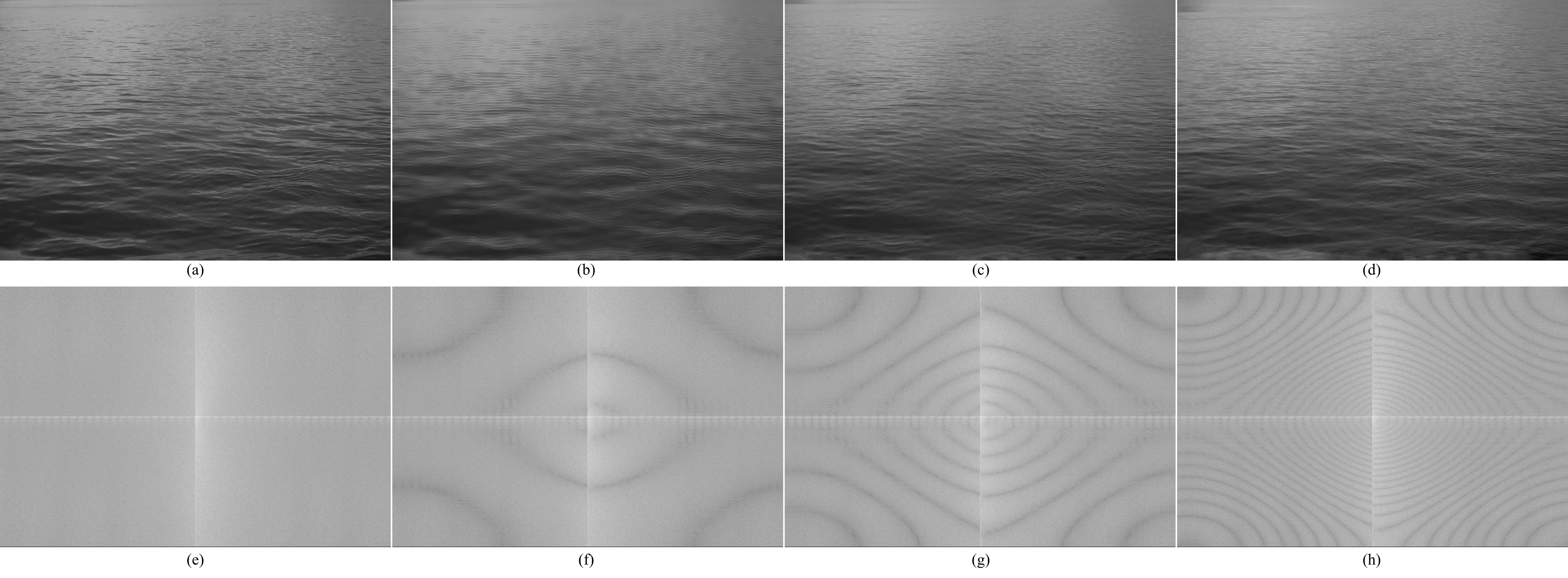}
    \caption{(a-d) Evolution of image and (e-h) Fourier spectrum under the wave equation.}
    \label{fig:Figure12}
\end{figure*}

In Fig. \ref{fig:Figure12}, one can observe the evolution of both the image and its Fourier spectrum (magnitude) driven by the wave equation. As the process progresses, the spectrum shows an increasing number of concentric circles, which visually highlight how the frequencies shift and propagate together through the image domain, much like waves.

Apart from being used to simulate waves propagating through pixels over an image, the wave equation is naturally equipped with the flexibility to restrict effects to regions defined by an optional mask. The mask, which can be generated using edge detection techniques and refined through post-processing (like morphological operations), allows selective wave dynamics, leaving unmasked regions static. This approach can be utilized for applications such as edge enhancement, segmentation, or various other tasks, depending on the problem to be solved.

\begin{figure*}[p!]
    \centering
    \includegraphics[width=\linewidth]{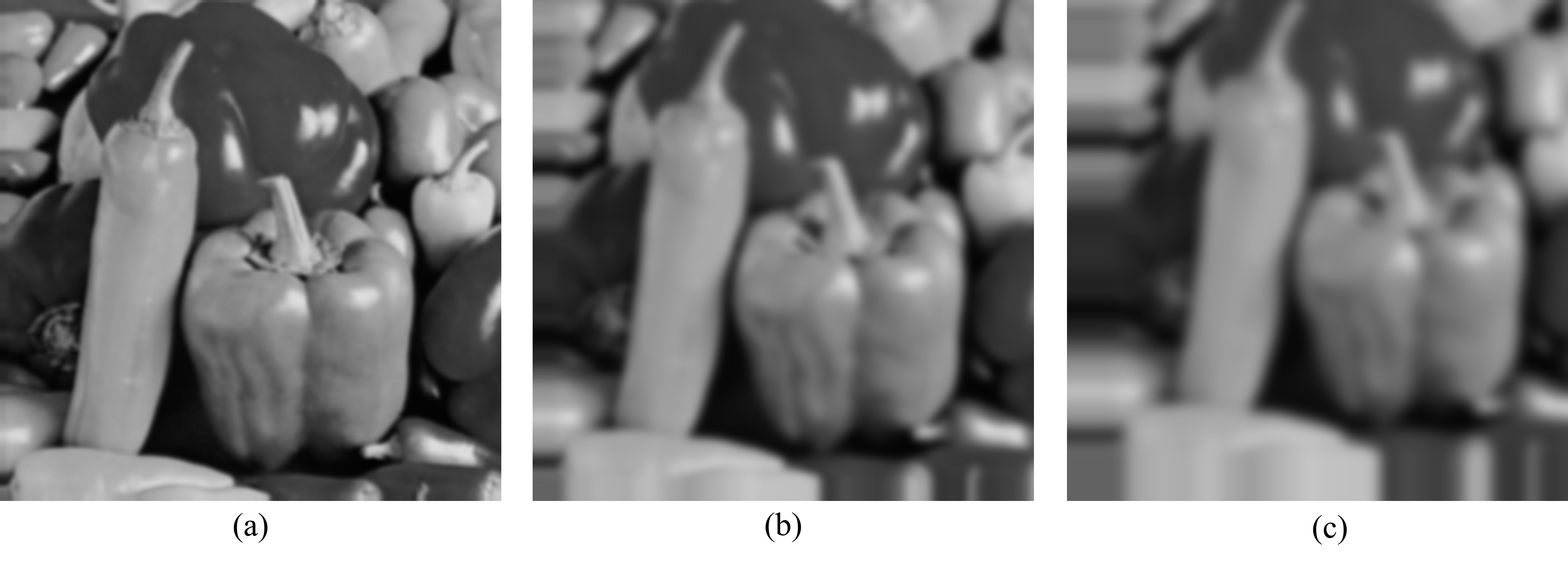}
    \caption{(a-c) Peppers image under the transport equation at three different time instances.}
    \label{fig:Figure13}
\end{figure*}

As can be seen in Fig. \ref{fig:Figure13}, the transport equation enables image registration and stabilization of cameras while simultaneously performing denoising through diffusion. It enables a precise real-time adjustment of the motion direction of the image, making it suitable for adaptive image control in video applications, such as surveillance or industry, to focus on a target and object tracking, also using a multiscale approach.

\begin{figure*}[p!]
    \centering
    \includegraphics[width=\linewidth]{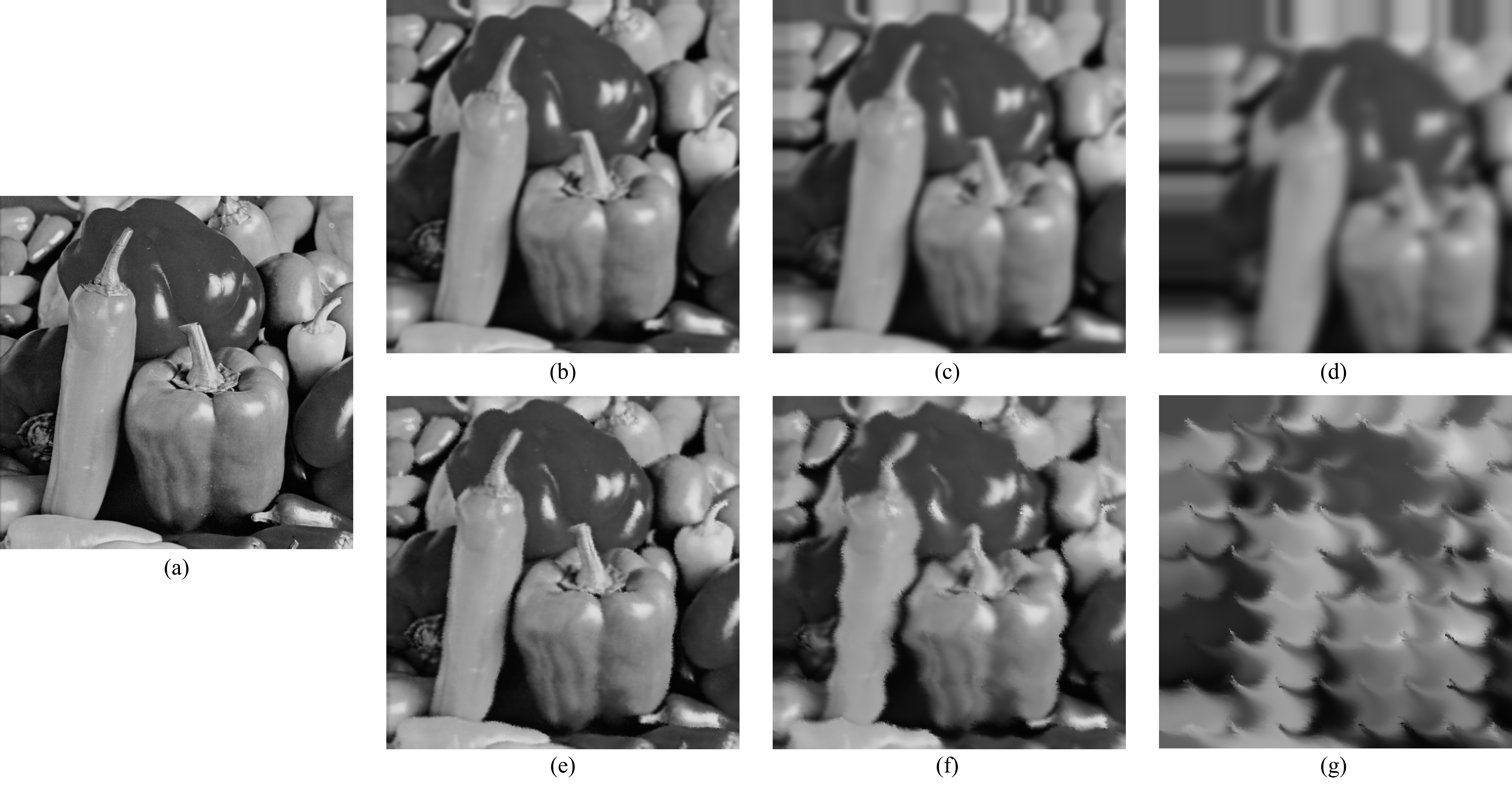}
    \caption{(a) Peppers image, under the Liouville equation at three different time instances, governed by (b-d) a constant velocity field and (e-g) a sinusoidal random one.}
    \label{fig:Figure14}
\end{figure*}

Moreover, Liouville equation can also be especially useful for non-rigid transformations during the registration process. Moreover, the diffusion incorporated into these equations allows for a multiscale effect, opening the door to complex studies in this application of PDEs (see Fig. \ref{fig:Figure14}).

\begin{figure*}[p!]
    \centering
    \includegraphics[width=\linewidth]{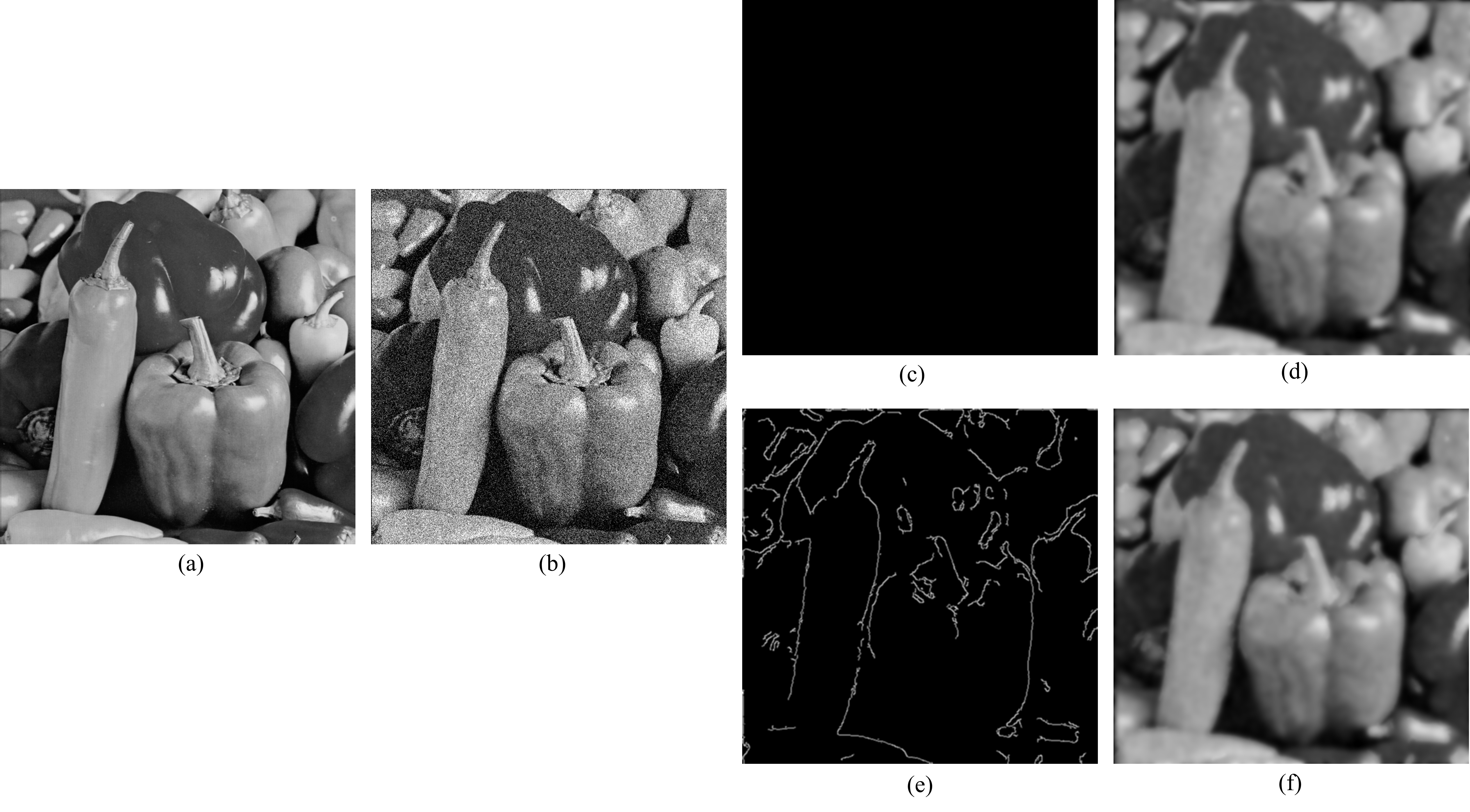}
    \caption{Results of Poisson’s PDE, (a) Peppers image, (b) noisy version with  , (c, d) mask and denoised result with no source, (e, f) mask and denoised result with a source.}
    \label{fig:Figure15}
\end{figure*}

Although similar to the heat and Laplace equations, the Poisson equation explains what it means for there to be a source of intensity, drawing an analogy between a charge or mass that creates a field (like an electric or gravitational field) and influences the surrounding space, and a source of intensity that represents regions of high contrast or significant gradient changes (which we wish to preserve). Hence a process of influence by the surrounding pixels can be posed as the image evolves by the Poisson PDE, as the exemplary result shown in Fig. \ref{fig:Figure15}. In can be seen that the object’s contours are accentuated as an effect of the intensity source integrated into the Poisson PDE, which, during the dynamic evolution of the image, counteracts the diffusive effect that typically degrades other heat-type equations.

\section{Conclusions}
The presented approaches utilized for denoising leverage the directional properties of intensity to enhance the robustness of noise removal techniques.
We have reviewed well-established techniques in PDE-based image processing and, building on these, introduced novel physics-based PDE models specifically designed for various image processing tasks. In particular, we have addressed challenges such as image denoising, deblurring, enhancement, and inpainting, presenting innovative models such as the Wave, Transport, Cahn-Hilliard, Burger, Liouville, Korteweg-de Vries, Kuramoto-Sivashinsky, and Maxwell-Heaviside equations.
We have experimentally demonstrated the performance of these methods, obtaining promising results; our findings show that these methods are applicable to a wide range of image processing tasks. Theoretically, we have laid the foundation for future investigations of the numerical resolution of these equations applied to images and other types of digital signals.
In the future, further studies are expected to explore the possibility of concatenating or integrating stages of PDEs with variational methods or investigating hybrid approaches and efficient solutions for complex functionals using Split-Bregman-type techniques, ADMM, or primal-dual approaches. Additionally, there is potential for the integration of fractional calculus techniques to take advantage of the growing field of convex optimization, which has become highly relevant in advanced image processing. The rigorous reformulation and proposal of the suggested PDEs from other perspectives is also highly encouraged.
The viscous Burgers equation seems to perform poorly according to the metrics due to the edge transport caused by the physical law it represents. However, it behaves much better in the boundaries of the domain, in contrast to other methods.
The adapted Maxwell equation has proven to perform very well in image restoration, particularly in inpainting, where it has outperformed all other methods considered in all tests. This makes it a very promising PDE that deserves further study from a physics-based perspective. Even the Maxwell-Heaviside PDE is capable of handling edges in a way that no other method can, as can be seen in the presented results.

\section{Data Availability}
The data used to support the findings of this study are available from the corresponding author upon request. The images were taken from \cite{Levkine2024}.

\section{Conflicts of interest}
The author declares that he has no known competing ﬁnancial interests or personal relationships that could have appeared to inﬂuence the work reported in this article.

\section{Acknowledgements}
This work did not receive financial support.

\bibliographystyle{ieeetr}
\typeout{}
\bibliography{export.bib}

\end{document}